\documentclass[12pt]{article}
\usepackage{amsfonts,amssymb,epsfig,amsmath}
\usepackage{color}

\usepackage[vcentermath]{youngtab}

\renewcommand{\baselinestretch}{1.2}

\begin{document}

\makeatletter \@addtoreset{equation}{section} \makeatother
\renewcommand{\theequation}{\thesection.\arabic{equation}}
\renewcommand{\thefootnote}{\alph{footnote}}

\begin{titlepage}

\begin{center}
\hfill {\tt SNUTP10-008}\\

\vspace{2.5cm}

{\Large\bf Holography of mass-deformed M2-branes}


\vspace{2.5cm}

\renewcommand{\thefootnote}{\alph{footnote}}

{\large Sangmo Cheon, Hee-Cheol Kim and Seok Kim}

\vspace{1cm}

\textit{Department of Physics and Astronomy \& Center for
Theoretical Physics,\\
Seoul National University, Seoul 151-747, Korea.}\\

\vspace{0.7cm}

E-mails: {\tt sangmocheon@gmail.com, heecheol1@gmail.com, skim@phya.snu.ac.kr}

\end{center}

\vspace{2.5cm}

\begin{abstract}

We find and study the gravity duals of the supersymmetric vacua of $\mathcal{N}\!=\!6$
mass-deformed Chern-Simons-matter theory for M2-branes. The classical solution extends
that of Lin, Lunin and Maldacena by introducing a $\mathbb{Z}_k$ quotient and discrete
torsions. The gravity vacua perfectly map to the recently identified supersymmetric
field theory vacua. We calculate the masses of BPS charged particles in the weakly
coupled field theory, which agree with the classical open membrane analysis when both
calculations are reliable. We also comment on how non-relativistic conformal symmetry
is realized in our gravity duals in a non-geometric way.

\end{abstract}

\end{titlepage}

\renewcommand{\thefootnote}{\arabic{footnote}}

\setcounter{footnote}{0}

\renewcommand{\baselinestretch}{1}

\tableofcontents

\renewcommand{\baselinestretch}{1.2}

\section{Introduction}

With recent advance in M2-brane physics from Chern-Simons-matter theories
\cite{Bagger:2006sk,Aharony:2008ug}, it became clear that many essential aspects
of M2-branes can be understood only when we have good controls over strongly
coupled quantum field theories. The strong coupling physics is crucial in
supersymmetry enhancement and appearance of M-theory states from monopole operators \cite{Aharony:2008ug,Kim:2009wb,Benna:2009xd,Gustavsson:2009pm,Kwon:2009ar,Bashkirov:2010kz},
the partition function and Wilson loops on $S^3$ \cite{Kapustin:2009kz}, determination of
exact $U(1)$ R-symmetry in $\mathcal{N}\!=\!2$ theories \cite{Jafferis:2010un}, the $N^{3/2}$
degrees of freedom \cite{Klebanov:1996un,Drukker:2010nc,Herzog:2010hf}, to list a few examples.
Analogous phenomena are either absent or turn out to be substantially
simpler in 4 dimensional Yang-Mills theories for D3-branes. Also, the physics often depends on
the Chern-Simons level $k$ in a more nontrivial way than the Yang-Mills coupling constant.

The roles of strong coupling dynamics turn out to be even more
important for understanding M2-brane systems with mass gap. The simplest M2-brane theory
with a mass gap is the mass-deformed $\mathcal{N}\!=\!6$ Chern-Simons-matter theory
\cite{Hosomichi:2008jb}. This theory has many discrete supersymmetric vacua, whose classical
solutions are first found in \cite{Gomis:2008vc} and refined in \cite{Kim:2010mr}.
When Chern-Simons level $k$ is $1$, the gravity
duals of the supersymmetric vacua are found by Lin, Lunin and Maldacena \cite{Lin:2004nb}.
See also \cite{Bena:2004jw}. A puzzle was that the number of the gravity solutions
is much smaller than that of the classical field theory vacua found in \cite{Gomis:2008vc}.
This puzzle was recently resolved in \cite{Kim:2010mr}. Many classically supersymmetric
vacua of \cite{Gomis:2008vc,Kim:2010mr} dynamically break supersymmetry, after which one
obtains a perfect agreement between the supersymmetric vacua of gauge theory and gravity
at $k\!=\!1$.\footnote{It
might sound surprising that the $\mathcal{N}\!=\!6$ theory admits dynamical supersymmetry
breaking. The analysis of \cite{Kim:2010mr} is from the
`UV completion' of this theory (Yang-Mills Chern-Simons-matter theory) which can have
no more than $\mathcal{N}\!=\!3$ supersymmetry. The result remains the same as one
flows to IR, continuously taking the Yang-Mills mass scale back to infinity.}

The analysis of \cite{Kim:2010mr} provides the partition function
(or more precisely, the index) of supersymmetric vacua at arbitrary Chern-Simons
level $k$, whose gravity duals are not well understood. The goal of this paper
is to identify and study the gravity duals of mass-deformed $\mathcal{N}\!=\!6$
theory at general coupling $k$.

Our results are very simple. The gravity duals for general $k$ can be obtained
from those for $k\!=\!1$ \cite{Lin:2004nb} by introducing $\mathbb{Z}_k$ quotients,
similar to the conformal Chern-Simons-matter theory \cite{Aharony:2008ug}. This fact
has been already noticed and partially studied in \cite{Auzzi:2009es}. Another important
aspect is that, contrary to the orbifold of $AdS_4\times S^7$, the orbifold of the
mass-deformed geometry has fixed points of the local form $\mathbb{R}^8/\mathbb{Z}_k$.
To correctly understand the gravity duals, one has to take into account the degrees of
freedom localized on these fixed points. We find that fractional M2-branes \cite{Aharony:2008gk}
are stuck to some of these fixed points, which appear in the gravity solutions
as discrete torsions. We find a set of gravity vacua which are in 1-to-1 map to the
field theory vacua of \cite{Kim:2010mr}, showing correct properties to be the gravity
duals of the latter.

Having obtained the precise map between the gravity backgrounds and the field theory
vacua, it could be possible to precisely address many questions on the gauge/gravity
duality of this system. For instance, the vortex solitons \cite{Kim:2009ny,Auzzi:2009es}
in this theory have been studied from the gravity duals \cite{Auzzi:2009es} using the
probe D0-brane analysis at large $k$. The results of this paper may help resolve
some of the puzzles concerning these objects, raised in \cite{Auzzi:2009es}.

M2-brane systems with mass-gap could also have potential applications to low
dimensional condensed matter systems. For instance, it is well known that quantum Hall
systems admit a low energy description based on Chern-Simons theory. Addition of
quasi-particles to this system would yield a Chern-Simons theory coupled to massive
charged matters. There are some studies of (fractional) quantum Hall systems based on
conformal Chern-Simons-matter theories \cite{Fujita:2009kw}. As the quantum
Hall systems are gapped in the bulk, it should be interesting to see if one can refine
these studies with the mass-deformed M2-brane systems.

Having such future directions aside, we study some basic properties of various
vacua, such as the spectrum of elementary excitations. After some part of the gauge
symmetry $U(N)\times U(N)$ is Higgsed in a vacuum, the unbroken part of the gauge group
sometimes exhibits non-perturbative dynamics, similar to the confinement in $\mathcal{N}\!=\!1$
Yang-Mills Chern-Simons theories studied in \cite{Witten:1999ds,Maldacena:2001pb}. This fact
can be captured by recently studied Seiberg-like dualities in 3 dimensions
\cite{Aharony:2008gk}, and also is correctly encoded in our gravity dual.
We also study the gravity duals of massive charged particles, given by open membranes
connecting various fractional M2-branes at the orbifold fixed points. The perturbative field
theory analysis of BPS charged particles is reliable when the 't Hooft couplings of unbroken
gauge groups are small, while classical open M2-brane analysis is reliable when the membrane is
macroscopic. We find a good agreement between the two spectra
when both calculations are reliable. Although we only discuss BPS particles
in this paper, similar comparison can be made with non-BPS particles.

The remaining part of this paper is organized as follows. In section 2, we first explain
the results of \cite{Kim:2010mr} which identifies the supersymmetric vacua of the
field theory. Then after explaining the gravity duals of the field theory vacua at
$k\!=\!1$, we introduce a $\mathbb{Z}_k$ orbifold and discrete torsions
for general Chern-Simons level $k$. Taking the fractional M2-branes into account,
we show that the supersymmetric ground states of the gravity solutions perfectly map to
the field theory vacua, with correct quantitative properties like
symmetry or M2-brane charges. The general considerations are illustrated by some
examples in section 3. In section 4, we study some elementary excitations in the
context of gauge-gravity duality. A brief comment on non-relativistic conformal symmetry
is also given. In section 5, we conclude with discussions on future directions.
Appendix A summarizes the supersymmetry of gravity solutions and various probe branes.
Appendix B explains the $SU(2)$ tensor description of the classical vacua. Appendix C
presents the mass calculation of charged particles from the field theory.

\section{The gravity soluions of mass-deformed M2-branes}

Chern-Simons-matter theory with mass deformation preserving
$\mathcal{N}\!=\!6$ Poincare supersymmetry was constructed and discussed in
\cite{Hosomichi:2008jb,Gomis:2008vc}. The theory has four complex scalars in
bi-fundamental representations of the $U(N)_k\times U(N)_{-k}$ gauge group, where
the subscripts $k$ and $-k$ denote the Chern-Simons levels. The classical and quantum
supersymmetric vacua of this theory have been studied in \cite{Gomis:2008vc} and \cite{Kim:2010mr},
respectively. At $k\!=\!1$, the gravity duals of the quantum supersymmetric vacua
are constructed in \cite{Lin:2004nb}, which are asymptotic to $AdS_4\times S^7$
in UV and exhibit complicated `bubbles' of M2-branes polarized into M5-branes
\cite{Bena:2000zb}. After semi-classical quantization of the 4-form fluxes, the
discretized gravity solutions are in 1-to-1 correspondence to the
supersymmetric vacua \cite{Kim:2010mr}.

In this section, we review the supersymmetric vacua of the field theory
identified in \cite{Kim:2010mr} for general $k$, explain the gravity duals for
$k\!=\!1$ obtained in \cite{Lin:2004nb}, and present the generalization for
arbitrary $k$. Then we propose the map between the gravity solutions and the field
theory vacua with various evidences.

\subsection{Supersymmetric vacua of the field theory}

Before mass deformation, there is an $SU(4)$ R-symmetry which rotates the
four complex scalars $Z_I$ (for $I=1,2,3,4$) in the fundamental representation.
The mass deformation breaks this R-symmetry to $SU(2)_1\times SU(2)_2\times U(1)$,
where $SU(2)_1$ and $SU(2)_2$ rotate $Z_1,Z_2$ and $Z_3,Z_4$ as doublets,
respectively. See \cite{Hosomichi:2008jb,Gomis:2008vc,Kim:2010mr} for the details.

\begin{figure}[t!]
  \begin{center}
    \includegraphics[width=8cm]{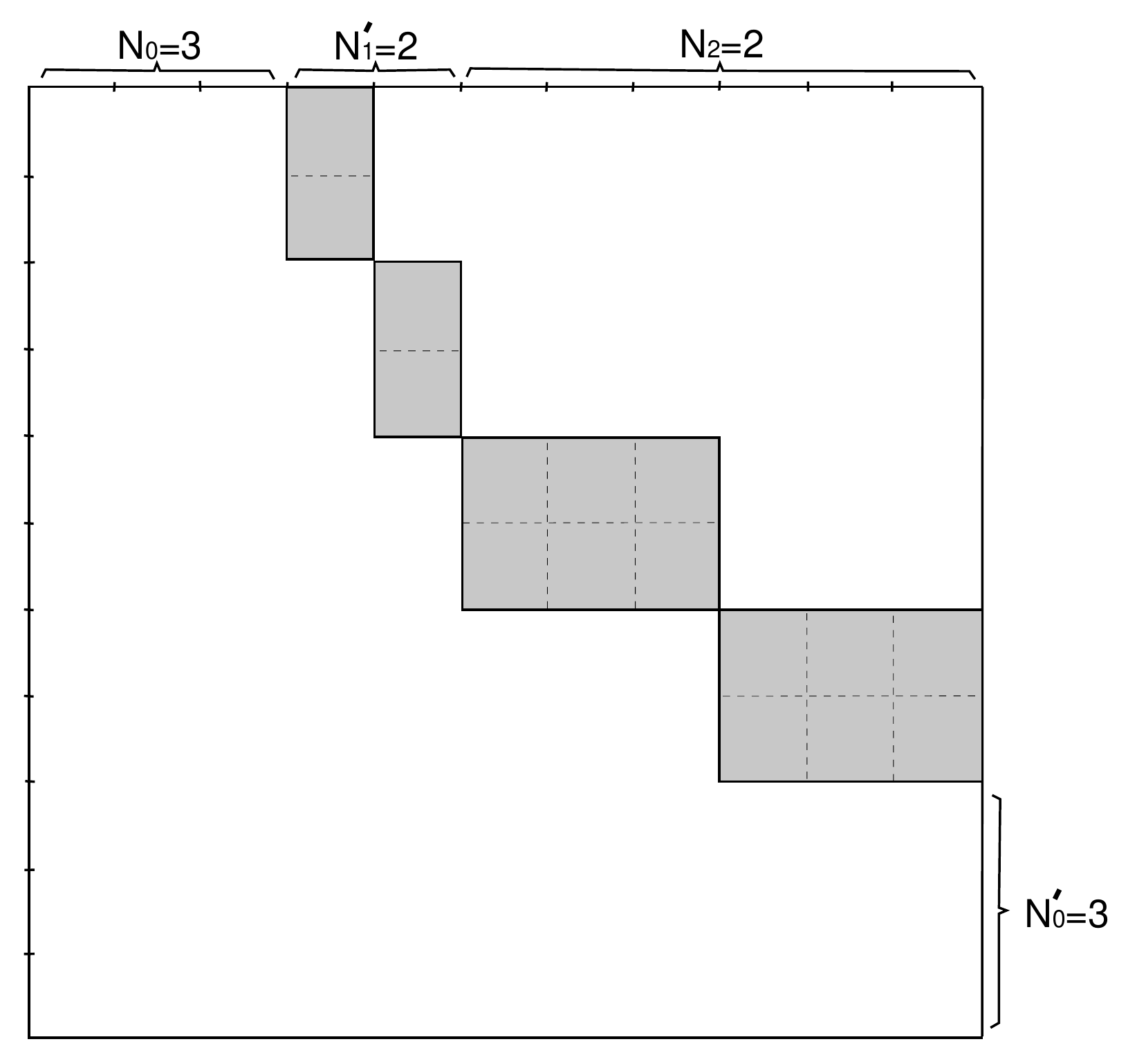}
\caption{An example of the parametrization of classical vacuum for $N\!=\!11$. Grey boxes
denote blocks with nonzero $Z_1,Z_2$ or $Z_3,Z_4$. In this figure, the
occupation numbers are $N_0\!=\!3$, $N_2\!=\!2$, $N_0^\prime\!=\!3$, $N_1^\prime\!=\!2$,
with all other numbers being zero.}\label{direct-sum}
  \end{center}
\end{figure}

The classical supersymmetric vacua are given by scalar configurations with
vanishing bosonic potential. The general solution is given by a direct sum of the
following blocks. The blocks of first type are $n\times(n\!+\!1)$
rectangular matrices with one more column, with $n\!=\!0,1,2,\cdots$. The blocks with
$n\!=\!0$ denote empty columns. In this block, two scalars $Z_3,Z_4$ are taken to be
zero, while the other two scalars are ($\mu$ is the mass parameter \cite{Kim:2010mr})
\begin{equation}\label{irreducible-1}
  Z_1=\mu^{\frac{1}{2}}\left(\begin{array}{cccccc}\sqrt{n}\!\!\!&0&&&&\\&\!\sqrt{n\!-\!1}
  \!\!&\!0&&&\\
  &&\ddots&\ddots&&\\&&&\sqrt{2}&0&\\&&&&1&0\end{array}\right),\
  Z_2=\mu^{\frac{1}{2}}\left(\begin{array}{cccccc}0&1&&&&\\&0&\sqrt{2}&&&\\
  &&\ddots&\ddots&&\\&&&0\!&\!\!\sqrt{n\!-\!1}\!&\\&&&&0&\!\!\!\sqrt{n}\end{array}\right).
\end{equation}
Below, we shall call this the $n$'th block of first type.
The blocks of second type are $(n\!+\!1)\times n$ rectangular matrices with
one more row, again with $n\!=\!0,1,2,\cdots$. Blocks with $n\!=\!0$ denote empty rows.
Here the two scalars $Z_1,Z_2$ are zero, while
\begin{equation}\label{irreducible-2}
  Z_3=\mu^{\frac{1}{2}}\left(\begin{array}{ccccc}\sqrt{n}\!\!\!&&&&\\0&\!\sqrt{n\!-\!1}&&&\\
  &0&\ddots&&\\&&\ddots&\sqrt{2}&\\&&&0&1\\&&&&0\end{array}\right)\ ,\ \
  Z_4=\mu^{\frac{1}{2}}\left(\begin{array}{ccccc}0&&&&\\1&0&&&\\
  &\sqrt{2}&\ddots&&\\&&\ddots&0&\\&&&\sqrt{n\!-\!1}\!&\!0\\&&&&\!\!\!\sqrt{n}\end{array}\right).
\end{equation}
We shall call this the $n$'th block of second type.
The classical supersymmetric vacua are parametrized by specifying how many blocks of different
types and sizes are included in the direct sum. We denote by $N_n$ the number of the $n$'th block
of first type, and by $N_n^\prime$ that of the $n$'th block of second type. An example
of our parametrization is shown in Fig \ref{direct-sum}.
As the direct sum (including $N_0^\prime$ empty rows and $N_0$ empty columns) should form an
$N\times N$ matrix, we obtain the following constraints on these `occupation numbers'
$\{N_n,N_n^\prime\}$:
\begin{equation}
  \sum_{n=0}^\infty\left[nN_n+(n\!+\!1)N_n^\prime\right]=N\ ,\ \
  \sum_{n=0}^\infty\left[(n\!+\!1)N_n+nN_n^\prime\right]=N\ .
\end{equation}
The first and second conditions are restrictions on the numbers of rows and columns.
Equivalently, we obtain the following two constraints
\begin{equation}\label{constraint-1}
  \sum_{n=0}^\infty\left(n\!+\!\frac{1}{2}\right)(N_n+N_n^\prime)=N\ ,\ \
  \sum_{n=0}^\infty N_n=\sum_{n=0}^\infty N_n^\prime\ ,
\end{equation}
which will be a more suggestive form for later interpretation.

Among these classical zero energy solutions, only some of them remain to be exactly
supersymmetric at the quantum level. We take the Chern-Simons level $k$ to be positive
without losing generality. The vacua which survive to be supersymmetric quantum
mechanically should satisfy \cite{Kim:2010mr}
\begin{equation}\label{constraint-2}
  0\leq N_n\leq k\ ,\ \ 0\leq N_n^\prime\leq k
\end{equation}
for all occupation numbers. Furthermore, when these restrictions
are satisfied, the degeneracy (or more precisely the Witten index) of the
supersymmetric vacua is
\begin{equation}\label{degeneracy}
  \prod_{n=1}^\infty\left(\begin{array}{c}k\\N_n\end{array}\right)
  \left(\begin{array}{c}k\\N_n^\prime\end{array}\right)\ ,
\end{equation}
for given $\{N_n,N_n^\prime\}$ \cite{Kim:2010mr}. The total number of supersymmetric
vacua with given rank $N$ (i.e. for a given theory) is the summation of the degeneracy
of the form (\ref{degeneracy}), taken for all occupation numbers satisfying
(\ref{constraint-1}) and (\ref{constraint-2}).

One can also generalize the identification of supersymmetric vacua in \cite{Kim:2010mr}
to the mass-deformed $\mathcal{N}\!=\!6$ theory with $U(N)_k\!\times\!U(N\!+\!\ell)_{-k}$
gauge group, where $0\leq\ell<k$ \cite{Aharony:2008gk}. The only change for the classical
supersymmetric vacua is to replace (\ref{constraint-1}) by
\begin{equation}
  \sum_{n=0}^\infty\left(n\!+\!\frac{1}{2}\right)(N_n+N_n^\prime)=N+\frac{\ell}{2}\ ,\ \ \sum_{n}(N_n-N_n^\prime)=\ell\ ,
\end{equation}
while the conditions (\ref{constraint-2}) and the degeneracy (\ref{degeneracy}) remain
the same.

\begin{figure}[t!]
  \begin{center}
    \includegraphics[width=8cm]{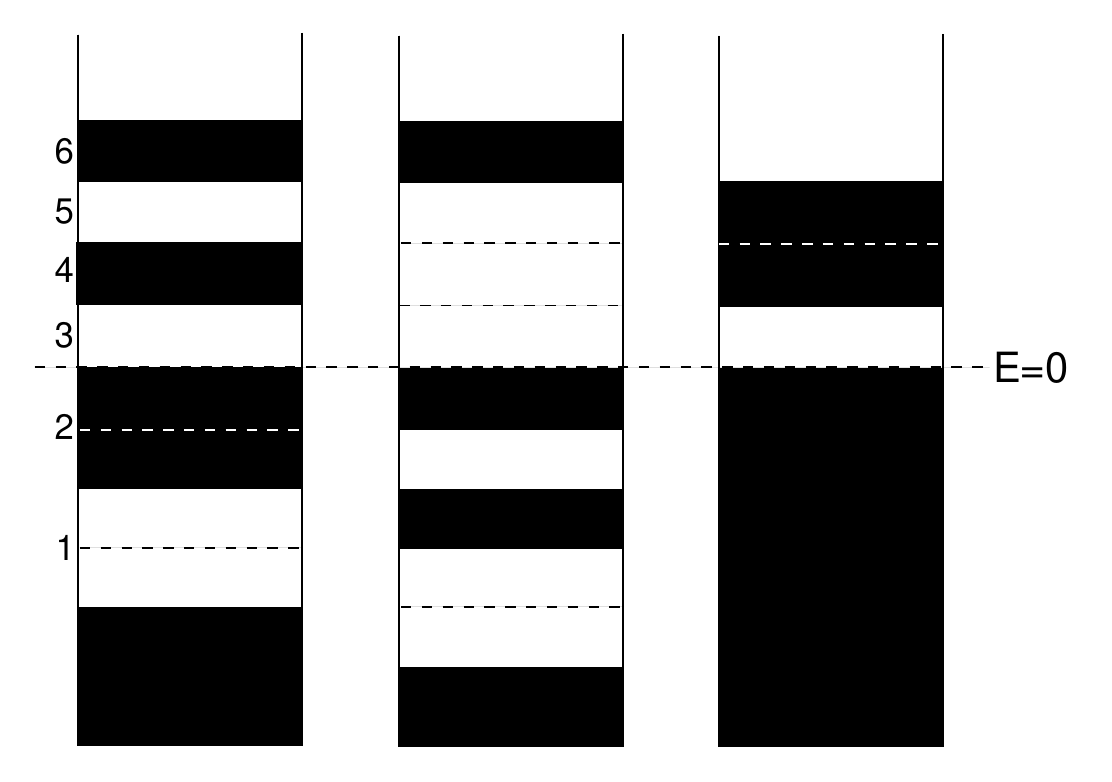}
    \includegraphics[width=8.5cm]{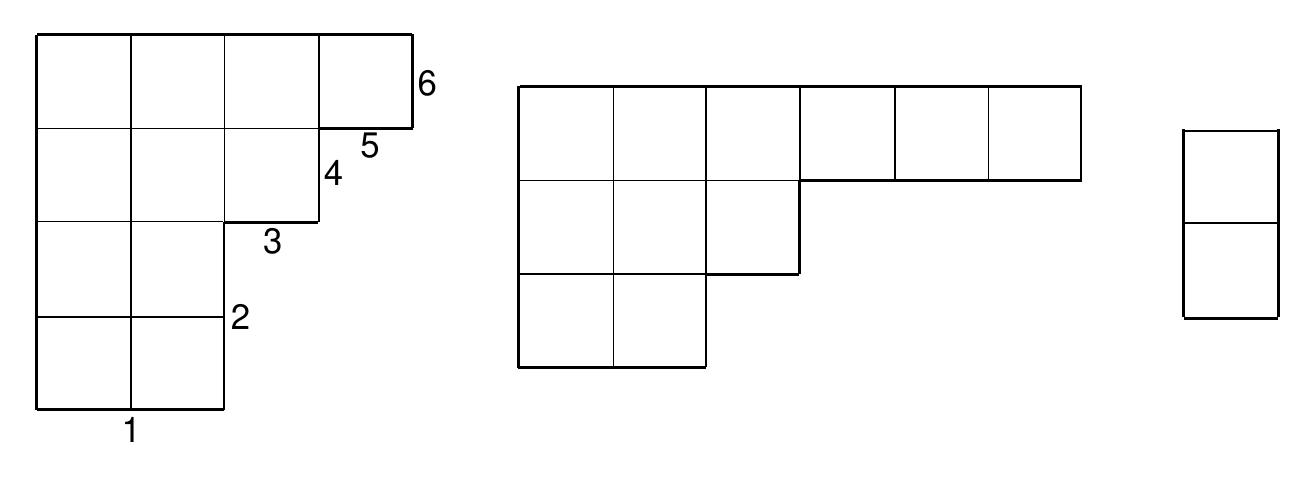}
\caption{A colored droplet for $k\!=\!3$ with $\ell\!=\!0$. Droplets have Fermi energies
$E_F=0,-2,2$ from left to right. The Young diagrams with charges $p\!=\!0,-2,2$ correspond
to the droplets: the edge lengths of the diagrams and the droplets match, as illustrated by
the numbers.}\label{colored-droplet}
  \end{center}
\end{figure}
Another way of viewing these supersymmetric vacua is to use $k$
species of fermions, or `colored' fermions, with occupation numbers $N_{ni}$, $N_{ni}^\prime$ (for $i=1,2,\cdots,k$) being either $0$ or $1$. Being blind to the $k$
`color' quantum numbers, one obtains the combinatoric factor (\ref{degeneracy}) in the degeneracy
by taking $N_n\!=\!\sum_{i=1}^kN_{ni}$, $N_n^\prime\!=\!\sum_{i=1}^kN_{ni}^\prime$.
The first condition of
(\ref{constraint-1}) about the rank $N$ is simply the overall energy condition for $k$
pairs of chiral fermions in $1\!+\!1$ dimensions. As the energy level $n+\frac{1}{2}$ is
half-integral, the fermions are in the Neveu-Schwarz sector. The second condition
of (\ref{constraint-1}) becomes the overall $U(1)$ singlet condition for the sum over the
$U(1)^{k}$ color charges, where the particles with the occupation numbers $N_{ni}$
carry charge $+1$ and anti-particles with occupation numbers $N_{ni}^\prime$ carry
charge $-1$. In other words, this condition sets the sum of $k$ Fermi energy levels to be $0$.
For the theory with nonzero $\ell$, the last condition modifies to setting
the sum of Fermi levels to be $\ell$. Such fermion viewpoint of the occupations can be illustrated
as `droplets' like Fig.\ref{colored-droplet}. In each droplet, an occupied sector with
$N_{ni}\!=\!1$ or $N_{ni}^\prime\!=\!1$ is represented by filling the $n$'th level above/beneath
the Fermi level
(denoted by $E\!=\!0$) with a black/white stripe, respectively. In this example, the field theory
occupation numbers are $N_4^\prime\!=\!1$, $N_3^\prime\!=\!2$, $N_2^\prime\!=\!1$, $N_1^\prime\!=\!1$,
$N_1\!=\!2$, $N_2\!=\!1$, $N_3\!=\!2$, and others zero.

One can also bosonize the $k$ colored fermions to $k$ chiral bosons on a circle,
namely to $k$ compact bosons.
The $U(1)^{k}$ charges become the quantized momenta $p_i$ (for $i\!=\!1,2,\cdots,k$)
of $k$ bosons on the circle. The neutrality of
overall $U(1)$ (or sum of Fermi levels being $\ell$) corresponds to the restriction
\begin{equation}\label{constraint-3}
  p_1+p_2+\cdots+p_k=\ell\ .
\end{equation}
The possible excitations of the $k$ bosons map to the so-called colored partitions of
$N$, which consists of $k$ Young diagrams with $U(1)^{k}$ charges $\{p_i\}$.
The first condition of (\ref{constraint-1}) on the energy of $k$ fermions can be
rewritten in the bosonized picture as
\begin{equation}\label{bosonized-energy}
  N+\frac{\ell}{2}=\frac{1}{2}\sum_{i=1}^{k}p_i^2+\sum_{i=1}^{k}\sum_{n=1}^\infty nN_{ni}\ ,
\end{equation}
where $N_{ni}\!=\!0,1,2,\cdots$ are excitations of the \textit{bosonic oscillators}
(not to be confused with fermionic occupations with same notation above), and $\{p_i\}$ are
subject to the constraint (\ref{constraint-3}). Note that the first term on the right hand side
of (\ref{bosonized-energy}) is the kinetic energy of $k$ zero modes.

In the bosonized picture, it is easy to calculate the partition function $Z(q)$ for the
supersymmetric vacua, which trades the energy with the chemical potential $q$ as
$Z(q)={\rm Tr}[q^{N\!+\!\frac{\ell}{2}}]$.
The partition function for the $k$ compact bosons with net momentum $\ell$ is
\begin{equation}
  I_{k}(q)=\prod_{n=1}^{\infty}\frac{1}{(1-q^n)^{k}}\sum_{p_1\!+\!p_2\!+\!\cdots\!+\!p_{k}
  \!=\ell}q^{\frac{1}{2}\sum_{i=1}^{k}p_i^2}\ .
\end{equation}
The first factor comes from the $k$ Young diagrams, while the second factor is from
the kinetic energy of zero modes. This is a generalization of the result of
\cite{Kim:2010mr} for $\ell\!=\!0$. For $k\!=\!1$,
$I_1(q)\!=\!\prod_{n=1}^\infty\frac{1}{1\!-\!q^n}$ agrees with the degeneracy of
the gravity solutions of \cite{Lin:2004nb}, as shown in \cite{Kim:2010mr}.
Strictly speaking, the above vacua are obtained by deforming the theory appropriately,
under which only the Witten index is invariant. So this partition function is the
Witten index of the field theory.

\subsection{Gravity solutions}

The gravity solutions for the supersymmetric mass-deformed M2-branes at Chern-Simons
level $k\!=\!1$, all asymptotic to $AdS_4\times S^7$, are obtained in \cite{Lin:2004nb}.
The metric and the 4-form field are given by \footnote{The 4-form flux $G_4$ corrects
the expression in \cite{Lin:2004nb},
$$
  \left[G_4\right]_{LLM}=
  -d\left(e^{2\Phi} h^{-2}V\right)\wedge dt\wedge dw_1\wedge dw_2-\frac{1}{4}
  e^{-2\Phi}\left[e^{-3G}\star_2d(y^2e^{2G})\wedge d\tilde\Omega_3+e^{3G}\star_2
  d(y^2e^{-2G})\wedge d\Omega_3\right]\ .
$$
which we think should contain typos. In particular, we explicitly checked
that the latter 4-form is not closed.}
\begin{eqnarray}\label{llm}
  ds^2&=&e^{\frac{4\Phi}{3}}\left(-dt^2+dw_1^2+dw_2^2\right)+
  e^{-\frac{2\Phi}{3}}\left[h^2(dy^2+dx^2)+ye^Gd\Omega_3^2+ye^{-G}
  d\tilde\Omega_3^2\right]\nonumber\\
  e^{-2\Phi}&=&\mu_0^{-2}\left[h^2-h^{-2}V^2\right]\nonumber\\
  G_4&=&-d\left(e^{2\Phi}h^{-2}V\right)\wedge dt\wedge dw_1\wedge dw_2
  +\mu_0^{-1}\left[Vd(y^2e^{-2G})-h^2e^{-3G}\star_2 d(y^2 e^{2G})\right]
  \wedge d\tilde\Omega_3\nonumber\\
  &&+\mu_0^{-1}\left[Vd(y^2e^{2G})+h^2e^{3G}\star_2 d(y^2 e^{-2G})\right]
  \wedge d\Omega_3\ ,
\end{eqnarray}
where $d\Omega_3$, $d\tilde\Omega_3$ denote length elements or volume 3-forms of unit
round 3-spheres, which we call $S^3$, $\tilde{S}^3$. Various functions in the
solution are determined by two functions $z(x,y)$, $V(x,y)$,
\begin{equation}\label{gravity-functions}
  z(x,y)=\sum_{i=1}^{2n\!+\!1}\frac{(-1)^{i\!+\!1}(x\!-\!x_i)}
  {2\sqrt{(x\!-\!x_i)^2+y^2}}\ ,\ \ V(x,y)=\sum_{i=1}^{2n\!+\!1}
  \frac{(-1)^{i\!+\!1}}{2\sqrt{(x\!-\!x_i)^2+y^2}}\ ,\ \
  ydV=-\star_2\ dz\ \ \ (\epsilon_{yx}=1)\ .
\end{equation}
The functions $G$ and $h$ are given by $z=\frac{1}{2}\tanh G$,
$h^{-2}=2y\cosh G$. Note that, compared to the solutions presented in
\cite{Lin:2004nb}, a parameter $\mu_0$ with dimension of mass is restored.
This could be eliminated to, say $\mu_0\!=\!1$, by using the asymptotic conformal
symmetry. $\mu_0$ will be identified with the mass paremeter $\mu$ appearing in the
field theory \cite{Kim:2010mr} as $\mu_0=\frac{\pi\mu}{2k}$.

\begin{figure}[t!]
  \begin{center}
    \includegraphics[width=10cm]{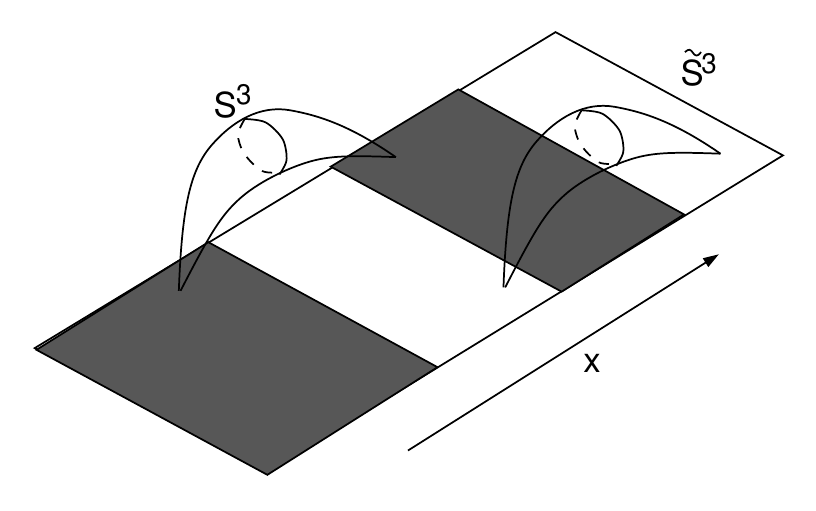}
\caption{The black/white regions with boundary conditions $z(x,y)\!=\!\mp\frac{1}{2}$.
The 4-sphere on the left combines a segment ending on black regions with $S^3$
shrinking at the ends. The second type of 4-sphere ends on white regions, and contains $\tilde{S}^3$.}\label{droplet-cycle}
  \end{center}
\end{figure}
From the metric in (\ref{llm}), $y^2$ is proportional to the product of the
square-radii of $S^3$ and $\tilde{S}^3$. Therefore, at least one of
the two 3-spheres shrink at $y\!=\!0$. For the geometry to be smooth with a
shrinking 3-sphere, the 3-sphere should combine with the radial direction
($\sim y$) to form $\mathbb{R}^4$. This requires the function $z$ in
(\ref{gravity-functions}) to have the boundary condition $z(x,0)=\mp\frac{1}{2}$,
where $S^3$ or $\tilde{S}^3$ shrinks for $\mp$ sign, respectively \cite{Lin:2004nb}.
At the line parametrized by $x$ at $y\!=\!0$, we therefore denote the parts with boundary
behaviors $z=\mp\frac{1}{2}$ by black/white regions, as shown in Fig \ref{droplet-cycle}.
To visualize the regions better, we add a fictitious line segment to make the $x$ line
look like an infinite strip of `droplet.' In type IIB dual, this extra segment has the meaning
of a spatial direction called $x^-$ in \cite{Lin:2004nb}, which is T-dualized to one of
the spatial coordinates of $\mathbb{R}^{2,1}$ in (\ref{llm}). To have asymptotic
$AdS_4\!\times\!S^7$, one should have a semi-infinite black region at one end and a white
region at the other end. At the boundary of the adjacent black and white regions
(call it $x\!=\!x_i$ for $i=1,2,\cdots 2n\!+\!1$), both 3-spheres shrink and $\mathbb{R}^8$
appears near $y\!=\!0$, $x\!=\!x_i$ by combining
the two 3-spheres with $x,y$.

There are various topological 4-cycles in this solution. Consider first a segment
in the $xy$ plane ending on different black regions at $y\!=\!0$, and attach the 3-sphere
$S^3$ to it, like the cycle on the left side of Fig \ref{droplet-cycle}. As $S^3$ shrinks
at the ends of the segment, the 4-cycle smoothly wraps up, forming
a 4-sphere. Similarly, one can consider a segment ending on different white
regions at $y\!=\!0$ and attach $\tilde{S}^3$ to it, which also becomes a 4-sphere as
shown by the cycle on the right side of Fig \ref{droplet-cycle}.
Nonzero 4-form fluxes are applied through these 4-spheres, which have to be quantized.
Below, we explain this quantization directly in M-theory. Similar discussion was provided
in \cite{Lin:2004nb} from the type IIB duals.

Consider a 4-sphere containing $\tilde{S}^3$ which surrounds a black region
($z\!=\!-\frac{1}{2}$) between $x\!=\!x_{2j}$ and $x_{2j\!+\!1}$,
where $j=0,1,\cdots,n$: see the cycle on the right side of Fig \ref{droplet-cycle}.
As the 4-form field is closed, we can deform the 4-sphere without changing the 4-form flux
over the cycle. We take the two points of the 4-sphere at $y\!=\!0$ to end exactly
at the boundaries of the black region. We also deform the whole 4-sphere to $y\!=\!0$.
Near this black region, with small $y$, one obtains
\begin{equation}
  z(x,y)\approx-\frac{1}{2}+\frac{y^2}{4}
  \left[-\sum_{i=1}^{2j}\frac{(-1)^{i\!+\!1}}{(x\!-\!x_i)^2}\!+\!
  \sum_{i=2j\!+\!1}^{2n\!+\!1}\frac{(-1)^{i\!+\!1}}{(x\!-\!x_i)^2}\right]
  \equiv-\frac{1}{2}+\frac{y^2}{4}\tilde{f}(x),\
  V(x,y)\approx\sum_i\frac{(-1)^{i\!+\!1}}{2|x\!-\!x_i|}\equiv
  \frac{1}{2}\tilde{g}(x)\ .
\end{equation}
Other functions are given by
\begin{equation}
  e^{2G}\approx\frac{y^2}{4}\tilde{f}(x)\ ,\ \
  h^{-2} \approx ye^{-G}\approx 2\tilde{f}^{-1/2}(x)\ ,\ \
  e^{-2\Phi}=h^2-h^{-2}V^2\approx\frac{1}{2\tilde{f}^{1/2}}\left(\tilde{f}
  -\tilde{g}^2\right)\ .
\end{equation}
Here, from
\begin{equation}
  \tilde{g}(x)=\sum_{i=1}^{2j}\frac{(-1)^{i\!+\!1}}{x\!-\!x_i}-
  \sum_{i=2j\!+\!1}^{2n\!+\!1}\frac{(-1)^{i\!+\!1}}{x\!-\!x_i}\ ,
\end{equation}
one finds that
\begin{equation}
  \tilde{g}^\prime=\tilde{f}\ .
\end{equation}
One can easily check that $\tilde{f}\!>\!0$ for $x_{2j}\!<\!x\!<\!x_{2j\!+\!1}$,
implying that $\tilde{g}$ is an increasing function
there. The case with $j\!=\!0$ (with $x_0\!=\!-\infty$) can be regarded as the semi-infinite
black region, in which only the upper boundary exists at $x\!=\!x_1$. The 4-form flux through
the 4-sphere surrounding $j$'th black region is given by
\begin{eqnarray}
  \mu_0\int G_4&=&2\pi^2
  \int_{2j}^{2j\!+\!1}\left[Vd\left(y^2 e^{-2G}\right)-h^2e^{-3G}\star_2
  d(y^2e^{2G})\right]=2\pi^2\int\left[2\tilde{g}d\left(\tilde{f}^{-1}\right)
  -\frac{1}{y^3\tilde{f}}\star_2 d\left(y^4\tilde{f}\right)\right]\nonumber\\
  &=&4\pi^2\int_{x_{2j}}^{x_{2j\!+\!1}}dx\left[\tilde{g}(\tilde{f}^{-1})^\prime+2\right]
  =4\pi^2\left[\left.\frac{}{}\tilde{g}\tilde{f}^{-1}\right|_{x_{2j}}^{x_{2j\!+\!1}}
  +\left(x_{2j\!+\!1}-x_{2j}\right)\right]\ ,
\end{eqnarray}
where we integrated by parts at the last step, using
$\tilde{g}^\prime=\tilde{f}$. For $j\geq 1$, one finds that the boundary contribution
from the first term is zero, and the flux is proportional to the `length'
$x_{2j\!+\!1}\!-\!x_{2j}$ of the black strip. This is basically the result of
\cite{Lin:2004nb} from the type IIB dual. For $j\!=\!0$, the boundary term
from $x\!=\!x_0\!\equiv\!-\infty$ is nontrivial. We temporarily set $x_0$ large but
finite as a regulator. $\tilde{g}\tilde{f}^{-1}$ contribution there is expanded as
\begin{equation}
  -\tilde{g}(x_0)\tilde{f}(x_0)^{-1}\approx x_0-\sum_{i=1}^{2n+1}(-1)^{i+1}x_i
  +\mathcal{O}(x_0^{-1})\ .
\end{equation}
Thus, the flux through this non-compact 4-cycle is given by
\begin{equation}\label{black-noncompact}
  \mu_0\int G_4=-4\pi^2\left[(x_{2n\!+\!1}\!-\!x_{2n})+
  (x_{2n\!-\!1}\!-\!x_{2n\!-\!2})+\cdots+(x_3\!-\!x_2)\right]\ ,
\end{equation}
$-4\pi^2$ times the total lengths of all finite black regions.
The type IIB picture \cite{Lin:2004nb} that the flux is proportional
to the area (or length in our case) of the black region is
true only for finite 4-cycles.

Similarly, one can calculate the flux through 4-spheres containing $S^3$
which surround white regions. On the white region between $x_{2j\!-\!1}$
and $x_{2j}$ ($1\!\leq\!j\!\leq\!n\!+\!1$, where the last entry is the
semi-infinite white region), the functions are expanded near $y\!=\!0$ as
\begin{eqnarray}
  z(x,y)&\approx&\frac{1}{2}-\frac{y^2}{4}
  \left[\sum_{i=1}^{2j\!-\!1}\frac{(-1)^{i\!+\!1}}{(x\!-\!x^i)^2}-
  \sum_{i=2j}^{2n\!+\!1}\frac{(-1)^{i\!+\!1}}{(x\!-\!x^i)^2}\right]
  \equiv\frac{1}{2}-\frac{y^2}{4}f\ ,\nonumber\\
  V(x,y)&=&\sum_i\frac{(-1)^{i\!+\!1}}{2|x\!-\!x^i|}\equiv
  \frac{1}{2}g(x)\ ,
\end{eqnarray}
where the functions now satisfy $g^\prime\!=\!-f$. Other functions are
given by $e^{-2G}\approx\frac{y^2}{4}f$, $h^{-2}\approx ye^G\approx
2f^{-1/2}$. The flux is given by
\begin{equation}
  \mu_0\int_{2j\!-\!1}^{2j}G_4=2\pi^2\int\left[Vd(y^2e^{2G})
  +h^2e^{3G}\star_2d(y^2e^{-2G})\right]=4\pi^2\left[
  \left.\frac{}{}gf^{-1}\right|_{x_{2j\!-\!1}}^{x_{2j}}-
  (x_{2j}\!-\!x_{2j\!-\!1})\right]\ .
\end{equation}
For $1\leq j\leq n$ with finite white regions, the boundary terms are zero
so that minus of the length of the strip gives the flux. For the semi-infinite
white region with $j\!=\!n\!+\!1$, one obtains
\begin{equation}
  \mu_0\int_{2n\!+\!1}^\infty G_4=4\pi^2\left[(x_{2n}\!-\!x_{2n\!-\!1})+
  (x_{2n\!-\!2}\!-\!x_{2n\!-\!3})+\cdots+(x_2\!-\!x_1)\right]\ ,
\end{equation}
which is the total length of the finite white strips.

The quantization condition of the 4-form fluxes on the 4-spheres is
\begin{equation}
  \frac{1}{(2\pi\ell_p)^3}\int_{S^4}G_4\in\mathbb{Z}\ ,
\end{equation}
where $\ell_p$ is the Planck length. The lengths of finite black/white regions
are thus quantized as
\begin{equation}\label{quantization}
  \frac{4\pi^2\mu_0^{-1}}{(2\pi\ell_p)^3}(x_{i\!+\!1}\!-\!x_i)\in\mathbb{Z}\ .
\end{equation}
In terms of the M2-brane tension $\tau_{M2}=\frac{2\pi}{(2\pi\ell_p)^3}$, one obtains
\begin{equation}
  x_{i\!+\!1}-x_i\in\frac{\mu_0}{2\pi\tau_{M2}}\mathbb{Z}=
  \frac{\mu}{4k\tau_{M2}}\mathbb{Z}\ .
\end{equation}
We shall later use the rescaled coordinates ${\bf x}_i$ of the boundaries
\begin{equation}
  {\bf x}_i=\frac{4\pi^2\mu_0^{-1}}{(2\pi\ell_p)^3}\ x_i\ .
\end{equation}
The distances between all boundaries are integers in the last coordinate.

\subsection{Discrete torsions and fractional M2-branes}

So far we have reviewed the solutions dual to $k\!=\!1$ vacua. For general $k$,
obvious solutions preserving the desired $\mathcal{N}\!=\!6$ supersymmetry are the
$\mathbb{Z}_k$ orbifolds of the above solutions. See appendix \ref{susy-gravity} for
the reduction of supersymmetry. We are not aware of an argument that such orbifolds
should give all $\mathcal{N}\!=\!6$ geometries. Here we simply assume this fact and
show in later sections that they are sufficient to understand aspects of dual field
theory (e.g. vacua, symmetry, elementary excitations), which strongly implies
that orbifolds of (\ref{llm}) are enough.

To take the $\mathbb{Z}_k$ orbifold, we consider the Hopf fibrations of $S^3$,
$\tilde{S}^3$, and take the two $U(1)$ angles
of the fibers to be $4\pi$ periodic $\psi$, $\tilde\psi$. The orbifold acts as
\begin{equation}
  \psi\rightarrow\psi+\frac{4\pi}{k}\ ,\ \ \tilde\psi\rightarrow
  \tilde\psi+\frac{4\pi}{k}\ .
\end{equation}
In the asymptotic $AdS_4\times S^7$ region, in which $S^3$ and $\tilde{S}^3$ combine
with one of the $x$, $y$ coordinates to form a round 7-sphere, this orbifold acts
freely on the Hopf fiber angle of $S^7$, which is the
known $\mathbb{Z}_k$ orbifold of $AdS_4\!\times\!S^7$ \cite{Aharony:2008ug}. However, the
orbifold has fixed points in the full mass-deformed geometry (\ref{llm}). Namely, near
$y\!=\!0$ and $x\!=\!x_i$ at the boundaries of black/white regions,
$S^3$ and $\tilde{S}^3$ combine to form $\mathbb{R}^8$, where $S^7$ can shrink. The
origin of $\mathbb{R}^8/\mathbb{Z}_k$ is a fixed point of the orbifold.

\begin{figure}[t!]
  \begin{center}
    \includegraphics[width=10cm]{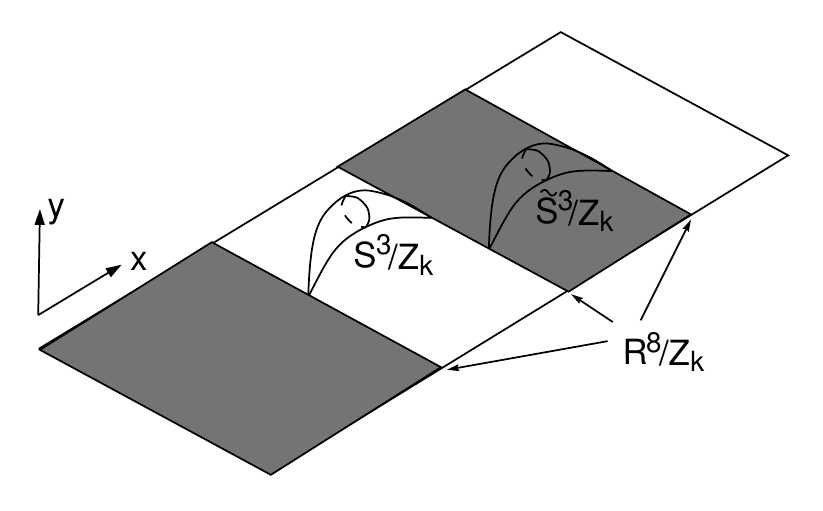}
\caption{The `4-cycles' $S^4/\mathbb{Z}_k$ and $\tilde{S}^4/\mathbb{Z}_k$ ending
on the edges of black/white regions, which have orbifold fixed points of the local form
$\mathbb{R}^4/\mathbb{Z}_k$ at the `north' and `south' poles.}\label{quotient-cycle}
  \end{center}
\end{figure}

With $\mathbb{Z}_k$ quotient, let us reconsider the 4-form fluxes
through various 4-cycles. Considering the covering space of $\mathbb{Z}_k$, the flux
quantization over 4-cycles still requires (\ref{quantization}) for the droplet.
When the two ends of the segment for the 4-cycle are placed exactly at the boundaries
of black/white regions as shown in Fig \ref{quotient-cycle},
one can also consider the 4-cycle which wraps $S^4/\mathbb{Z}_k$ only. Such `4-cycles'
have two orbifold singularities of the form $\mathbb{R}^4/\mathbb{Z}_k$ inherited from
$\mathbb{R}^8/\mathbb{Z}_k$ at the two ends, which we call north and south poles
of $S^4$. The 4-form flux through this $S^4/\mathbb{Z}_k$ is an integer divided by $k$, which
is fractional in general. This fractional flux through $S^4/\mathbb{Z}_k$ can be consistent
with the flux quantization as one can turn on nonzero
\textit{discrete torsions} of the 3-form field $C_3$ at the two $\mathbb{R}^8/\mathbb{Z}_k$
\cite{Aharony:2008gk}. Discrete torsion on $S^7/\mathbb{Z}_k$ is given by the holonomy of $C_3$
on the torsion 3-cycle $S^3/\mathbb{Z}_k$,
\begin{equation}\label{torsion-fw-anomaly}
  \frac{1}{(2\pi\ell_p)^3}\int_{S^3/\mathbb{Z}_k}C_3=-\frac{\ell}{k}+\frac{1}{2}
\end{equation}
where $\ell$ is an integer which can be put in, say, $0\leq\ell<k$ by a gauge
transformation $\ell\sim\ell+k$. The shift $\frac{1}{2}$ is recently claimed to be important
\cite{Aharony:2009fc} due to the so-called Freed-Witten anomaly, which is also consistent
with the large gauge transformation of Page charges. $-k$ times this expression, $\ell\!-\!\frac{k}{2}$,
is the `Page charge' for the M5-branes wrapping a collapsing 3-cycle in $\mathbb{R}^8/\mathbb{Z}_k$ \cite{Aharony:2009fc}. One can turn on such a discrete torsion at each
$\mathbb{R}^8/\mathbb{Z}_k$ fixed point. Let us call $\ell_i$ the discrete torsion
at the fixed point $x=x_i$. The discrete torsions $\ell_i$ and $\ell_{i\!+\!1}$
on the adjacent fixed points in $x$ direction are related, due to the
flux conditions on $S^4/\mathbb{Z}_k$ cycles just mentioned. The flux on the orbifolded
4-cycle stretched between $x\!=\!x_i$ and $x_{i\!+\!1}$ is
\begin{equation}
  \frac{1}{(2\pi\ell_p)^3}\int_{S^4/\mathbb{Z}_k}G_4=
  -\frac{1}{(2\pi\ell_p)^3}\int_{S^3/\mathbb{Z}_k,\ {\rm north}}C_3
  -\frac{1}{(2\pi\ell_p)^3}\int_{S^3/\mathbb{Z}_k,\ {\rm south}}C_3\ ,
\end{equation}
where the integrals on the right hand sides are taken with the outgoing orientations
from the north/south poles. The left hand side is an integer divided by $k$. As the two
terms on the right hand sides are
torsions at $x\!=\!x_i$, $x_{i\!+\!1}$ up to integer shifts, we arrive at the recursion
relation between adjacent discrete torsions
\begin{equation}\label{recursion}
  \ell_i\!+\!\ell_{i\!+\!1}=\pm({\rm quantized\ length\ between}\
  x_i\sim x_{i\!+\!1})\ \mod k\ ,
\end{equation}
where the $\pm$ signs are for the black/white regions, respectively.
From these relations, all torsions on the fixed points can be decided
once we know the discrete torsion of the asymptotic
$AdS_4\times S^7/\mathbb{Z}_k$, either at $x\!=\!\pm\infty$, $y\!=\!0$.

The asymptotic discrete torsion at $x\!=\!\pm\infty$, which we call
$\ell\!=\!\ell_0=\ell_{2n\!+\!2}$, is related to the rank of the gauge group
in the dual field theory \cite{Aharony:2008gk}. Namely, taking the torsion
in $0\leq\ell<k$ with $k>0$, the field theory comes with $U(N)_k\times U(N\!+\!\ell)_{-k}$
group.\footnote{We work with the convention of \cite{Aharony:2009fc}, which is
different from \cite{Aharony:2008gk}, related to the sign choice in
(\ref{torsion-fw-anomaly}).}
Using (\ref{recursion}), one can start from
$x\!=\!x_0\!=\!-\infty$ at $y=0$ and proceed by increasing $x$, determining
$\ell_1,\ell_2,\cdots$ in turn.
As the recursion
formula (\ref{recursion}) is given in the `outgoing' convention from the edge,
we should change $\ell\rightarrow-\ell$ before applying this formula.
The torsion at $x_1$ is given by
\begin{equation}\label{1st-torsion}
  \ell_1=-({\bf x}_{2n\!+\!1}\!-\!{\bf x}_{2n})-
  ({\bf x}_{2n\!-\!1}\!-\!{\bf x}_{2n\!-\!2})-\cdots-({\bf x}_3\!-\!{\bf x}_2)
  +\ell\mod k\ .
\end{equation}
Next, since the flux on the first white region is $-({\bf x}_2\!-\!{\bf x}_1)$, one obtains
\begin{equation}
  \ell_2=-\ell_1-({\bf x}_2-{\bf x}_1)=\sum_{i=1}^{n}({\bf x}_{2i\!+\!1}\!-\!{\bf x}_{2i})
  -({\bf x}_2\!-\!{\bf x}_1)-\ell\mod k\ ,
\end{equation}
and so on. One can continue this analysis to obtain
\begin{eqnarray}
  \ell_{2i}&=&({\rm black})_{n}+({\rm black})_{n\!-\!1}+\cdots
  +({\rm black})_i-({\rm white})_1-\cdots-({\rm white})_i-\ell\nonumber\\
  \ell_{2i\!+\!1}&=&-({\rm black})_{n}-({\rm black})_{n\!-\!1}-\cdots
  -({\rm black})_{i\!+\!1}+({\rm white})_1+\cdots+({\rm white})_i+\ell\nonumber
\end{eqnarray}
modulo $k$, where (black)$_i$ and (white)$_i$ denote the
quantized lengths of the $i$'th black and white regions.
In particular, the torsion at $x_{2n\!+\!1}$ is given by
\begin{equation}
  \ell_{2n\!+\!1}=+({\rm white})_1+\cdots+({\rm white})_n+\ell\mod k\ .\nonumber
\end{equation}
From $\ell_{2n\!+\!1}\!+\!\ell_{2n\!+\!2}\sim\int_{2n\!+\!1}^\infty
G_4$, we find that the torsion at $x=\infty$ is
$\ell_{2n\!+\!2}\!=\!-\ell\mod k$ in the outgoing orientation from infinity
towards the droplet, agreeing with the torsion at $x=-\infty$.

In \cite{Aharony:2008gk}, it was argued that discrete torsions on
$\mathbb{R}^8/\mathbb{Z}_k$ can be interpreted as fractional M2-branes at the
orbifold. In our case, we should be careful to identify whether discrete torsions are
to be identified as fractional M2-branes or anti M2-branes. By anti M2-branes, we mean
those which carry negative M2-brane charges in the convention that the total
M2-charge of the gravity solution is positive. As the issue of orientation is
same for both full and fractional M2-branes, we consider full M2-branes to understand
the charge signs of fractional M2-branes.

\begin{figure}[t!]
  \begin{center}
    \includegraphics[width=7.9cm]{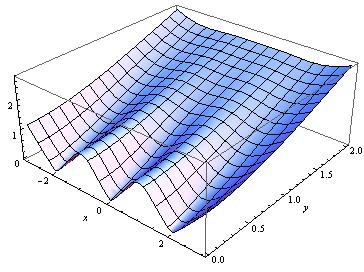}
    \includegraphics[width=7.9cm]{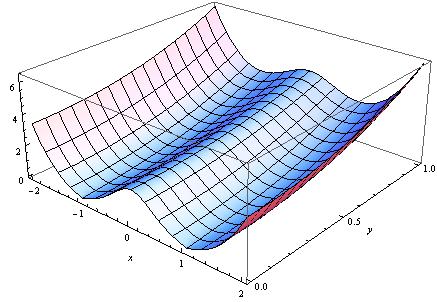}
\caption{The potential energy density for an M2-brane (left) and an anti M2-brane
(right) on the $xy$ space, in a background given by the droplet with edges at
${\bf x}_1\!=\!-2$, ${\bf x}_2\!=\!-1$, ${\bf x}_3\!=\!0$, ${\bf x}_4\!=\!1$,
${\bf x}_5\!=\!2$. [Plotted for $\mu_0\!=\!1$, with the $x$ axes being the
dimensionless coordinate ${\bf x}$.]}\label{probe-m2}
  \end{center}
\end{figure}
Let us first consider the dynamics of probe M2-branes
(full branes, not fractional) in the background (\ref{llm}). These M2-branes are extended
in $\mathbb{R}^{2,1}$ with
two possible orientations, and are transverse in the 8 dimensional space spanned by
$x,y$ and $S^3$, $\tilde{S}^3$. One obtains the following potential energy density from
the Nambu-Goto action and the Wess-Zumino coupling,
\begin{equation}\label{m2-potential}
  e^{2\Phi}\pm C_{012}=e^{2\Phi}\mp e^{2\Phi}h^{-2}V=\frac{1}{h^2\pm V}\ ,
\end{equation}
where the upper/lower signs are for the M2- and anti M2-branes, respectively.
The signs $\pm$ for M2- and anti M2-branes can be easily determined by demanding
the two contributions from $e^{2\Phi}$ and $C_{012}$ have more cancelation for
M2's than anti-M2's in the UV region. This should be the case since
the former has two contributions exactly canceling with each other in
$AdS_4\times S^7$ without mass deformation. From these potentials,
one finds that M2-branes stabilize at $y=0$ and $x=x_1,x_3,x_5,\cdots$, while
anti M2-branes stabilize at $y=0$ and $x=x_2,x_4,x_6,\cdots$. Fig \ref{probe-m2}
shows the potentials for M2- and anti M2-branes for the droplet with 5 edges.

The supersymmetry analysis of appendix \ref{susy-gravity} shows that \textit{both}
M2-branes (at odd edges $x_{2i\!+\!1}$) and anti M2-branes (at even edges $x_{2i}$)
are supersymmetric. This is because the supersymmetry condition of the gravity
solution reduces to $\Gamma^{012}\epsilon=\pm\epsilon$ with different signs at odd
and even edges, making M2's and anti-M2's supersymmetric there. Therefore,
we conclude that the fractional M2-branes at odd and even edges have same
orientations as the full branes there. The BPS M2-branes with negative charge have a
natural interpretation, on which we shall elaborate in section 3. As we shall illustrate
in more detail there, we claim that gravity solutions containing negatively charged
fractional M2-branes have to be excluded for comparing with the field theory vacua. A simple reasoning
is that (either full or fractional) anti M2-branes with negative M2-charge in a
background with positive charge can be geometrized to yield a solution which
contains only positively charged fractional M2-branes. Therefore, solutions containing
negatively charged M2-branes provide redundant descriptions of the field theory vacua.

Below in this subsection and also in section 2.4, we concentrate on the gravity
solutions containing positively charged fractional M2's only and show that they
perfectly map to the supersymmetric field theory vacua with correct physical properties.

At a fixed point with torsion $\ell$, the low energy effective description for these fractional
M2-branes is pure $U(\ell)_{-k}$ Chern-Simons theory \cite{Aharony:2008gk} with $0\leq\ell<k$,
in the convention of \cite{Aharony:2009fc} that we are advocating.
More precisely, this was derived with $\mathcal{N}\!=\!3$ supersymmetric UV theory
given by Yang-Mills Chern-Simons theory. In \cite{Aharony:2008gk}, this theory was argued
to be Seiberg-dual to $U(k\!-\!\ell)_{k}$ theory, by studying the D-brane realization of
this system. In particular, this implies that the $U(k)_{\pm k}$ theory is dual to nothing. This could
be closely related to the fact that $\mathcal{N}\!=\!1$ Yang-Mills Chern-Simons theory with
$SU(2k)_k$ \cite{Witten:1999ds} and $U(2k)_k$ \cite{Maldacena:2001pb} gauge groups are
confining. This aspect will be very important for us later when we try to map
the gravity solutions to the field theory, as the low energy description of a field theory
vacuum will also be given by Chern-Simons theories with products of $U(\ell)_{\pm k}$ like
groups. We should have in mind that, the field theory factors of $U(k)_{\pm k}$ would be dual
to nothing and thus absent in the gravity duals. Or oppositely, \textit{for the convenience of
comparing our gravity solutions to the field theory vacua}, we may work in the
gravity dual side with some formally assigned `fictitious $U(k)_{\pm k}$' Chern-Simons sectors
as they are dual to nothing: the meaning of the last point will be clear below.

\begin{figure}[t!]
  \begin{center}
    \includegraphics[width=11cm]{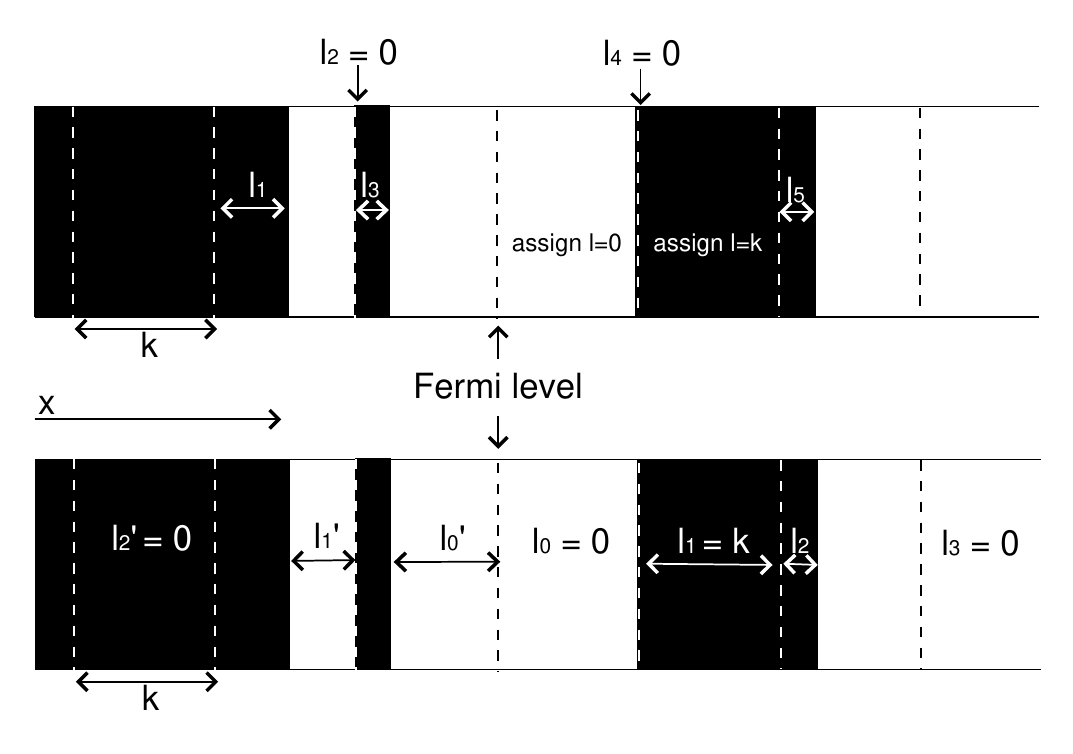}
\caption{Dotted lines divide the droplet into strips of length $k$.
In the upper droplet, nonzero torsions at $x\!=\!x_{2i\!+\!1}$ can be regarded as being
assigned to length $k$ strips, given by the lengths of the black region inside a strip.
Strips with only black or white region are formally assigned with torsion $\ell\!=\!0$ or $k$.
In the lower droplet, we Seiberg-dualize the Chern-Simons theory below the Fermi level:
torsions below the Fermi level are then lengths of white regions inside a strip.}\label{torsion}
  \end{center}
\end{figure}
The droplets with nonzero torsions only at the `odd edges' $x_{2i\!+\!1}$ have a simple
structure which will be useful later when comparing them to field theory.
To simplify the story, we restrict our interest in the remaining part of this paper to
the case in which the asymptotic torsion at $AdS_4\times S^7$ is zero, with gauge group
$U(N)\times U(N)$: the generalization to the case with $U(N)\times U(N\!+\!\ell)$ is
straightforward. We first note that, as even edges $x_{2i}$ support zero torsions, the quantized
distances between the even edges are all multiples of $k$. This is easily seen from
the recursion relation (\ref{recursion}),
\begin{equation}\label{k-strip}
  \ell_{2i}\!+\!\ell_{2i\!+\!1}=({\bf x}_{2i\!+\!1}-{\bf x}_{2i}) \mod k\ ,\ \
  \ell_{2i\!+\!1}\!+\!\ell_{2i\!+\!2}=-({\bf x}_{2i\!+\!2}-{\bf x}_{2i\!+\!1}) \mod k\
  \rightarrow\ {\bf x}_{2i\!+\!2}\!-\!{\bf x}_{2i}\in k\mathbb{Z}\ ,
\end{equation}
since all $\ell_{2i}$ are $0$. Therefore, we naturally divide the droplet into
strips of length $k$, as shown in the upper droplet of Fig \ref{torsion} (where
divisions are denoted by dashed lines). By construction,
all even edges are boundaries of these strips. There could also be some boundaries of
these strips which simply cut a long black or white regions into length $k$ strips.
In each strip of length $k$, there is a black strip of length $\ell$ on the left side
and a white strip of length $k\!-\!\ell$ on the right side, for $0\leq\ell\leq k$. The case
with $\ell\!=\!0$ or $k$ happens when the whole strip of length $k$ is just white or black.
When $\ell\!\neq\!0,k$ in a given length $k$ strip, $\ell$ is the torsion
at $x_{2i\!+\!1}$ or the number of fractional M2-branes, from (\ref{k-strip}).
They are described by $U(\ell)_{-k}$ Chern-Simons theory. When
$\ell\!=\!0$ or $k$, we do not have to assign any such degree, as the fractional
M2-branes are either absent or dual to nothing. Just for convenience, we also \textit{formally}
assign $U(\ell)_{-k}$ Chern-Simons theories even there: this will not affect the physics
anywhere, as long as one remembers that $U(k)_{-k}$ theory is dual to nothing. In this way,
we assign $U(\ell)_{-k}$ type Chern-Simons theories (or discrete torsion $\ell$) to all strips
of length $k$, not to the edges.

It is helpful to label the above strips of length $k$ by introducing
the notion of Fermi level, which is obtained by taking all black strips down to fill
the white regions below them, to get a droplet with one semi-infinite black and white
regions only. The $x$ location of the boundary between the two semi-infinite regions is
called the Fermi level. One can easily prove that the Fermi level is at one of the boundaries
of the above strips of length $k$, namely at a dashed line in Fig \ref{torsion}. To show
this, recall that the torsion at $x\!=\!x_1$ is minus of the
total quantized lengths of all black strips above $x_1$ (mod $k$),
given by (\ref{1st-torsion}). The distance from $x\!=\!x_1$
to the other end of the neighboring white region is ${\bf x}_2\!-\!{\bf x}_1$, so
\begin{equation}
  \ell_2+\ell_1=-({\bf x}_2-{\bf x}_1)
\end{equation}
from the relation between adjacent torsions. Since $\ell_2=0$, one finds that
\begin{equation}\label{lowest-torsion}
  \ell_1=-({\bf x}_2-{\bf x_1})=-({\rm total\ length\ of\ finite\ black\ strips})
  \ \mod k\ ,
\end{equation}
where the second equation is from (\ref{1st-torsion}).
Now trying to pull down all black regions to fill in the whites below them,
one first uses some of the black regions to fill the first white region of length
${\bf x}_2\!-\!{\bf x}_1$. The remaining total lengths of finite black regions that can
be pulled down is a multiple of $k$, from the second equation in (\ref{lowest-torsion}).
Thus, after these remaining blacks are pulled down, the height of the Fermi level is
a multiple of $k$ plus $x_2$, which is at the boundary of the above strips of length $k$,
as claimed.

Now we parametrize the strips of length $k$ from the Fermi level. The
strips above the Fermi level are labeled by non-negative integers $n\!=\!0,1,2,\cdots$,
where $n$ increases as we get farther from the Fermi level. The strips below the Fermi
level are labeled similarly, again by increasing non-negative integers
as we move down away from the Fermi level. In each strip, the length of the black region
contained in this strip is the discrete torsion assigned to this strip. Let us
relabel the discrete torsions by denoting by
$\ell_n$ the torsion in the $n$'th strip above the Fermi level, and by $\tilde\ell_n$
the torsion in the $n$'th strip below the Fermi level. When a strip
consists of black or white regions only, we call the corresponding torsion to be
$k$ or $0$, respectively: remember that both are equivalent to no fractional M2's.

At low energy, the fractional M2-branes in the bulk are described by pure Chern-Simons theory
(having $\mathcal{N}\!=\!3$ supersymmetric UV completion) with gauge group and level
\begin{equation}\label{gravity-original}
  \prod_{n=0}^\infty U(\ell_n)_{-k}\times U(\tilde\ell_n)_{-k}\ .
\end{equation}
One can also perform Seiberg duality transformations of \cite{Aharony:2008gk} to obtain
a different description, which will be more directly related to the
field theory later. Performing this
duality for all fractional branes below the Fermi level, and defining
${\ell_n}^\prime\equiv k\!-\!\tilde\ell_n$, one obtains a Chern-Simons theory with
\begin{equation}\label{gravity-symmetry}
  \prod_{n=0}^\infty U(\ell_n)_{-k}\times U({\ell_n}^\prime)_k\ .
\end{equation}
gauge group and levels. An example of such a parametrization is given in the
lower droplet of Fig \ref{torsion}. Below the Fermi level after Seiberg duality,
the torsions assigned to a strip of length $k$ is given by the length of the white
region in the strip. (\ref{gravity-symmetry}) takes the same form as the unbroken
gauge symmetry of the field theory side \cite{Kim:2010mr}, if one identifies
$\ell_n\!=\!N_n$, $\ell_n^\prime\!=\!N_n^\prime$. More will be addressed about this
map in the next subsection.

\subsection{M2 charge, symmetry and map to the field theory vacua}

Now we turn to the problem of mapping the gravity solutions to the field theory
vacua explained in section 2.1. In this subsection, we provide three supporting
evidences for our proposed map: symmetry, M2-brane charges, and the degeneracy of the
vacua for given $N,k$. In section 4, more evidence will be provided by comparing
the spectra of elementary excitations of field theory and open membranes in the
gravity duals.

We start by recalling that the low energy description
in a field theory vacuum with $N_n$ first type blocks of size $n$ and $N_n^\prime$ second
type blocks of size $n$ ($n\!=\!0,1,2,\cdots$) is given by the Chern-Simons theory with
gauge group and Chern-Simons levels \cite{Kim:2010mr}
\begin{equation}\label{unbroken}
  \prod_{n=0}^\infty U(N_n)_{-k}\times U(N_n^\prime)_k\ .
\end{equation}
Comparing this with the low energy Chern-Simons theory for fractional M2-branes in the
gravity dual with gauge group (\ref{gravity-symmetry}), one is naturally led to consider
the map
\begin{equation}\label{map}
  \ell_n=N_n\ ,\ \ \ell_n^\prime=N_n^\prime\ .
\end{equation}
As both parameters $\ell_n,\ell_n^\prime$ and $N_n,N_n^\prime$ range between $0$ and $k$,
(\ref{map}) provides a map between the classical gravity solution and a classical
field theory vacuum solution which do not dynamically break supersymmetry. With this map,
the gauge theory and gravity sides have same low energy descriptions with same symmetry
groups and Chern-Simons levels.

Obviously, as the low energy descriptions are same on both sides, the vacuum
degeneracy of the field theory and gravity precisely match with each other.
For each factor of $U(\ell)_{\pm k}$ Chern-Simons theory, we consider it with
a supersymmetric UV completion of an $\mathcal{N}\!=\!3$ supersymmetric Yang-Mills
Chern-Simons theory. The vacuum degeneracy of this Chern-Simons
theory is given by \cite{Witten:1999ds,Ohta:1999iv}
\begin{equation}
  \left(\begin{array}{c}k\\ \ell\end{array}\right)=\frac{k!}{\ell!(k\!-\!\ell)!}\ .
\end{equation}
Therefore, both degeneracies from gauge theory and gravity are given by
\begin{equation}
  \prod_{n=1}^\infty\left(\begin{array}{c}k\\N_n\end{array}\right)
  \left(\begin{array}{c}k\\N_n^\prime\end{array}\right)=
  \prod_{n=1}^\infty\left(\begin{array}{c}k\\ \ell_n\end{array}\right)
  \left(\begin{array}{c}k\\ \ell_n^\prime\end{array}\right)\ ,
\end{equation}
agreeing with each other.

As the gauge group rank $N$ is given in terms of $\{N_n,N_n^\prime\}$ by
(\ref{constraint-1}), and as $N_n,N_n^\prime$ are identified with the torsions
$\ell_n,\ell_n^\prime$ in the gravity solutions, one should also see if the M2-brane
number $N$ calculated from gravity is consistent with (\ref{constraint-1}).
Discrete torsions (or fractional M2-branes) at various orbifold fixed points
contribute to the M2-brane charge. This has been calculated rather recently in \cite{Bergman:2009zh,Aharony:2009fc}. We are interested in the Maxwell M2-brane charge,
\begin{equation}\label{maxwell-def}
  Q_{M2}=\frac{1}{(2\pi\ell_p)^6}\int_{\left[S^7/\mathbb{Z}_k\right]_\infty}\star\ G_4\ ,
\end{equation}
where $\left[S^7/\mathbb{Z}_k\right]_\infty$ denotes the 7-sphere in the
asymptotic region. If one computes this at asymptotic infinity, the calculation
is the same as that obtained in $AdS_4\times S^7$ without mass deformations.
The result is \cite{Aharony:2009fc} (see also \cite{Bergman:2009zh})
\begin{equation}\label{m2-charge-infty}
  Q_{M2}=\left(N+\frac{k}{8}\right)+b\left(\ell-\frac{k}{2}\right)+\frac{kb^2}{2}
  -\frac{1}{24}\left(k-\frac{1}{k}\right)=N+\frac{\ell(k-\ell)}{2k}
  -\frac{1}{24}\left(k-\frac{1}{k}\right)
\end{equation}
where $b\equiv-\frac{\ell}{k}+\frac{1}{2}$ is the $C_3$ torsion. $N+\frac{k}{8}$ appearing
in the first parenthesis is the M2-brane Page charge, with quantized $N$ \cite{Aharony:2009fc}.
In our case, with $\ell\!=\!0$, one finds that
\begin{equation}\label{m2-charge-uv}
  Q_{M2}=N-\frac{1}{24}\left(k-\frac{1}{k}\right)\ .
\end{equation}
The number $N$ is the rank of the gauge group $U(N)\!\times\!U(N)$ in the field theory.

On the other hand, in the full mass-deformed geometry, the M2-brane charge can also be
computed with the IR data, from the droplets and torsions. To find this expression,
we deform the integration 7-manifold in (\ref{maxwell-def}) to the IR region $y\approx 0$.
As the Maxwell charge is not localized due to the equation of motion for $C_3$,
\begin{equation}
  d\star G_4=-\frac{1}{2}\ G_4\wedge G_4\ ,
\end{equation}
one obtains
\begin{equation}\label{m2-charge-domain}
  Q_{M2}=\frac{1}{(2\pi\ell_p)^6}\left[\int_{\mathcal{D}_7}\star\ G_4-\frac{1}{2}
  \int_{\mathcal{M}_8}G_4\wedge G_4\right]\ .
\end{equation}
$\mathcal{D}_7$ is the shrinking region with $y\approx 0$,
and $\mathcal{M}_8$ is an 8 dimensional space spanned by $x,y$ and two 3-spheres,
which has $\partial\mathcal{M}_8\!=\![S^7/\mathbb{Z}_k]_\infty\!-\!\mathcal{D}_7$
as its boundary. For $k\!=\!1$, as the geometry is smooth, $\mathcal{D}_7$
is void and the charge can be calculated solely from the last term of
(\ref{m2-charge-domain}). For general $k$, $\mathcal{D}_7$ can be taken to be
the union of small $S^7/\mathbb{Z}_k$ regions surrounding the $\mathbb{R}^8/\mathbb{Z}_k$
singularities at the edges of black/white regions $x\!=\!x_i$, $y\!=\!0$.

The second contribution of (\ref{m2-charge-domain}) proportional to $G_4\wedge G_4$
can be computed from the covering space of our $\mathbb{Z}_k$ quotient solution, which
should be divided by $k$. From the Young diagram
picture of the droplet, this is
simply given by the number of boxes in the Young diagram. See our section 2.1 as well
as \cite{Lin:2004nb} for this conversion. From the `fermion droplet' picture, the
Young diagram size is the total `energy' of all particles (black region
above the Fermi level) and holes (white region below Fermi level) in the
NS sector, as explained in section 2.1.
For the droplet parametrized by $\ell_n,\ell_n^\prime$, this quantity is given by
\begin{equation}\label{m2-charge-1}
  kQ_{M2}\leftarrow\sum_{n=0}^\infty\left[\sum_{i_n=0}^{\ell_n\!-\!1}\left(
  kn\!+\!i_n\!+\!\frac{1}{2}\right)+\sum_{i_n=0}^{\ell_n^\prime\!-\!1}\left(kn\!+\!i_n\!+\!\frac{1}{2}
  \right)\right]=\sum_{n=0}^\infty\left[kn(\ell_n+\ell_n^\prime)+\frac{\ell_n^2+(\ell_n^\prime)^2}{2}
  \right]\ ,
\end{equation}
where $+\frac{1}{2}$ shifts in all parentheses appear as fermions are in the NS sector.

With $\mathbb{Z}_k$ fixed points, there appear extra contributions from the singularities
and discrete torsions contained in regions surrounded by $\mathcal{D}_7$.
These have been computed in \cite{Bergman:2009zh,Aharony:2009fc} for each factor of
$S^7/\mathbb{Z}_k$, which also take the form of (\ref{m2-charge-infty}).
Let us consider the contribution from a singularity at odd edges and even edges
in turn. At an odd edge $x\!=\!x_{2i\!+\!1}$, there are no full M2-branes. Combining
other contributions to $Q_{M2}$, one obtains from (\ref{m2-charge-infty})
\begin{equation}
  Q_{M2}^{(2i\!+\!1)}=\left(0+\frac{k}{8}\right)-\frac{1}{2k}\left(\ell_{2i\!+\!1}-\frac{k}{2}
  \right)^2-\frac{1}{24}\left(k-\frac{1}{k}\right)=\frac{\ell_{2i\!+\!1}(k\!-\!\ell_{2i\!+\!1})}{2k}
  -\frac{1}{24}\left(k-\frac{1}{k}\right)\ ,
\end{equation}
where $\ell_{2i\!+\!1}$ is the torsion ranged in $0\leq\ell_{2i\!+\!1}<k$.
On the other hand, at even edges $x_{2i}$, the assigned torsion is zero from the requirement
that no fractional anti M2-branes be there. So the only contribution is coming from
the curvature at the singularity \cite{Bergman:2009zh}. One can easily check that the curvature
contribution comes with a relative minus sign compared to the odd edges,
\begin{equation}
  Q_{M2}^{(2i)}=+\frac{1}{24}\left(k-\frac{1}{k}\right)\ .
\end{equation}
One way to see this is continuously deforming the droplet to place an even edge
on top of an adjacent odd edge. This can be done, ignoring the quantization of droplets,
as the curvature contribution to $Q_{M2}$ can be obtained solely from classical considerations.
As the two edges merge and annihilate, the curvature contributions from the two $\mathbb{Z}_k$
fixed points cancel, leading to the above sign flip. Summing over all contributions from
the singularities, one obtains
\begin{equation}
  \frac{1}{(2\pi\ell_p)^6}\int_{\mathcal{D}_7}\star\ G_4=Q_{M2}^{(1)}+Q_{M2}^{(2)}+\cdots+
  Q_{M2}^{(2n\!+\!1)}=\sum_{i=0}^n\frac{\ell_{2i\!+\!1}(k\!-\!\ell_{2i\!+\!1})}{2k}-
  \frac{1}{24}\left(k-\frac{1}{k}\right)\ .
\end{equation}
In our new parametrization of the droplet/torsions with reference to the
Fermi level, the torsions $\ell_{2i\!+\!1}$ are to be identified with either $\ell_n$ or
$\tilde\ell_n$ in our new parametrization introduced around (\ref{gravity-original}).
Although the latter set of torsions also include formally assigned extra torsions $0$ or $k$,
equivalent to nothing, they do not affect the expression in the summation as
$\frac{\ell(k\!-\!\ell)}{2k}$ is zero for both $\ell\!=\!0,k$.
Also, this expression does not change by shifting to a Seiberg-dual description, changing
$\tilde\ell_n$ to $\ell_n^\prime=k-\tilde\ell_n$. One thus obtains
\begin{equation}
  \frac{1}{(2\pi\ell_p)^6}\int_{\mathcal{D}_7}\star\ G_4=\sum_{n=0}^\infty
  \left[\frac{\ell_n(k\!-\!\ell_n)}{2k}+\frac{\ell_n^\prime(k\!-\!\ell_n^\prime)}{2k}\right]
  -\frac{1}{24}\left(k-\frac{1}{k}\right)\ .
\end{equation}
Adding this to (\ref{m2-charge-1}), one obtains
\begin{equation}
  Q_{M2}=\sum_{n=0}^\infty\left(n\!+\!\frac{1}{2}\right)
  \left(\ell_n+\ell_n^\prime\right)-\frac{1}{24}\left(k-\frac{1}{k}\right)\ .
\end{equation}
Comparing this with the expression (\ref{m2-charge-uv}) obtained from UV, one obtains
\begin{equation}\label{rank-map}
  N=\sum_{n=0}^\infty\left(n\!+\!\frac{1}{2}\right)(\ell_n+\ell_n^\prime)\ .
\end{equation}
This relation is exactly the same as the first equation in
(\ref{constraint-1}) for the field theory vacua, if we identify the torsions
$\{\ell_n,\ell_n^\prime\}$ with the field theory occupation numbers
$\{N_n,N_n^\prime\}$. We find this is a very nontrivial consistency check of
our proposal for the following reason. Our map was first based on the requirement
that two low energy descriptions agree. However, the constraint
from the same low energy descriptions is quite weak since one suffices to demand
$\{\ell_n\}\!=\!\{N_n\}$ and $\{\ell_n^\prime\}\!=\!\{N_n^\prime\}$ with arbitrarily
shuffled order. On the other hand, (\ref{rank-map})
demands that the subscripts of $N_n$ and $\ell_n$ be equal to have same $N$.

The `level matching condition,' the second equation in (\ref{constraint-1}),
can also be obtained from the gravity solution. Namely, $\ell_n$ is the length
of the black region in $n$'th strip of length $k$ above the Fermi level, while
$\ell_n^\prime$ is the length of the white region in $n$'th strip below Fermi level.
By definition of the Fermi level, the total length of the black regions above the
Fermi level equals that of the white regions below the Fermi level, leading to
\begin{equation}
  \sum_{n=0}^\infty\ell_n=\sum_{n=0}^\infty\ell_n^\prime\ .
\end{equation}
This agrees with the second condition of (\ref{constraint-1}).

We end this section with a comment on the unpolarized vacuum, where the classical
expectation values of all scalars are zero. In this case, the only nonzero occupation
numbers are $N_0=N$ and $N_0^\prime=N$. This vacuum breaks supersymmetry if $N>k$, namely
when the gravity dual has a chance to be weakly curved. On the other hand, for $N\leq k$,
the unpolarized vacuum for $N$ M2-branes remains supersymmetric and is included in our
gravity solutions. They necessarily have string scale curvature. Firstly in the UV
region, this is true as the radius of curvature in the string unit is proportional to
$\left(N/k\right)^{1/4}$ \cite{Aharony:2008ug}. One can also check that the IR radii
of curvatures in the $y\!=\!0$ region of various 2-spheres are small in the string scale.
Extending one's interest to vacua which spontaneously break supersymmetry, it should be
interesting to find and study the gravity solution for the unpolarized vacuum with
$N\gg k$, trying to address the gravity duals of nonrelativistic conformal theories.
See section 4.2 as well as the discussion section for more comments.

\section{Examples}

In this section, we provide concrete examples of the gravity solutions and their
map to the field theory vacua, to supplement the formal considerations in the previous
section. With various examples, we also argue why the gravity
solutions containing negatively charged fractional M2-branes should be regarded as
redundant solutions.

We start by noting that droplets can be equivalently described by Young diagrams,
as explained in section 2.1 with Fig \ref{colored-droplet}. The black/white regions
map to the vertical/horizontal edges of the Young diagram. We use the compact
Young diagram description for the illustration.

\begin{figure}[t!]
  \begin{center}
    \includegraphics[width=2.5cm]{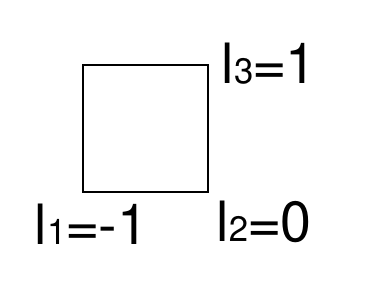}
\caption{Young diagram (=gravity solution) for $k\!=\!2$, $N\!=\!1$}\label{k2n1}
  \end{center}
  \begin{center}
    \includegraphics[width=10cm]{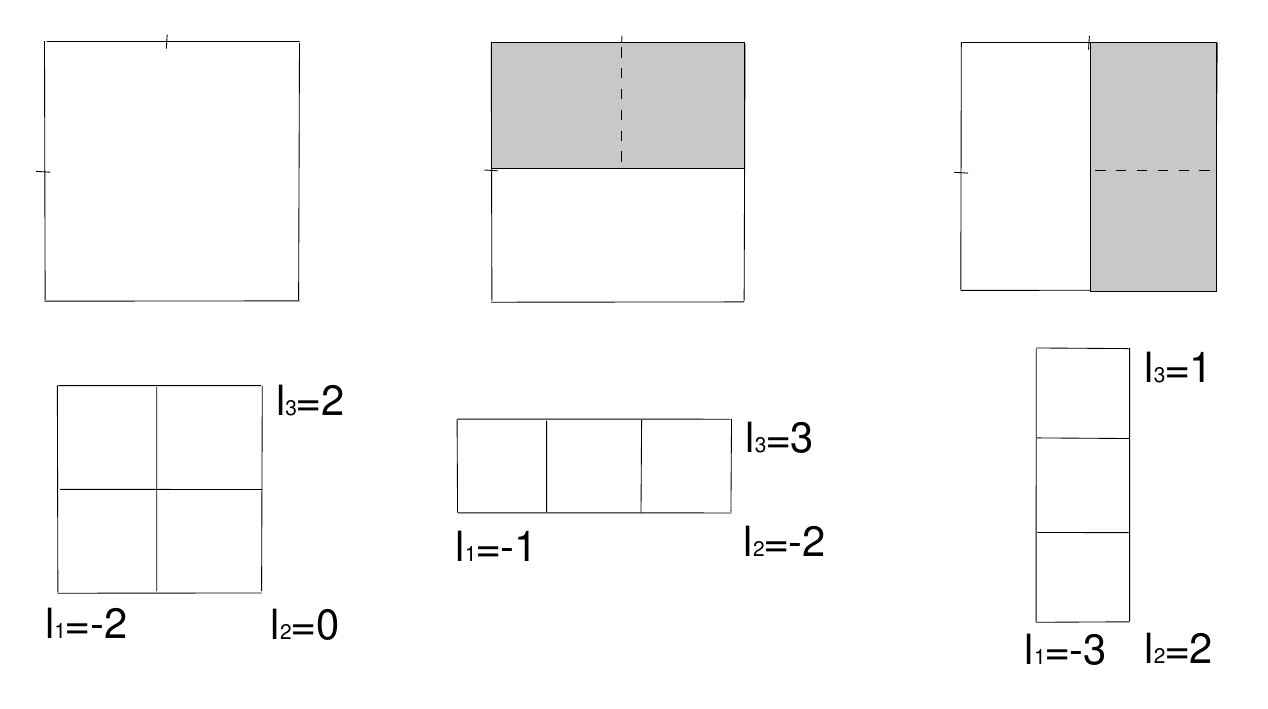}
\caption{$k\!=\!2$, $N\!=\!2$: $2\times 2$ matrices for supersymmetric vacua (upper figures).
Young diagrams without negatively charged fractional M2-branes (lower figures)}\label{k2n2}
  \end{center}
\end{figure}
The examples for $k\!=\!2$ with various M2-brane charge $N$ are given as
follows. Firstly, for $N\!=\!1$, there is only one field theory vacuum
$N_0\!=\!N_0^\prime\!=\!1$. The corresponding Young diagram is made of one box,
and its discrete torsion data is shown in Fig \ref{k2n1}. After
using $\ell\!\sim\!\ell\!+\!2$ to bring all torsions to be either $0$ or $1$, the odd
torsions carry M2-brane charge $\frac{1\cdot(k\!-\!1)}{2k}\!=\!\frac{1}{4}$,
while even torsions carry M2-brane charge $0$.
The numbers at the corners of the Young diagram are the discrete torsions
$\ell_1$, $\ell_2$, $\ell_3$ at the $\mathbb{R}^8/\mathbb{Z}_2$ singularities, which are
defined mod 2. These can be obtained from the recursion relation (\ref{recursion}).

For $N\!=\!2$, there are three classical supersymmetric vacua which survive to be
supersymmetric at the quantum level. The three vacua are given by $N_0\!=\!N_0^\prime\!=\!2$, $N_1\!=\!N_0^\prime\!=\!1$ and $N_0\!=\!N_1^\prime\!=\!1$. As scalar $2\times 2$ matrices,
they look like the upper figures in Fig \ref{k2n2}. The shaded boxes denote insertions of
nonzero blocks. The Young diagrams have various numbers of boxes, which together with
M2-charge from torsions should satisfy $N\!=\!2$. Nonzero discrete torsions at convex corners
of the Young diagram (i.e. even edges of the droplet) correspond to fractional anti M2-branes,
carrying negative M2-brane charges. The gravity solutions which do not have fractional
anti M2-branes are listed in the lower figures of Fig \ref{k2n2}:
each Young diagram map to the classical vacuum at the same column.
\begin{figure}[t!]
  \begin{center}
    \includegraphics[width=17.5cm]{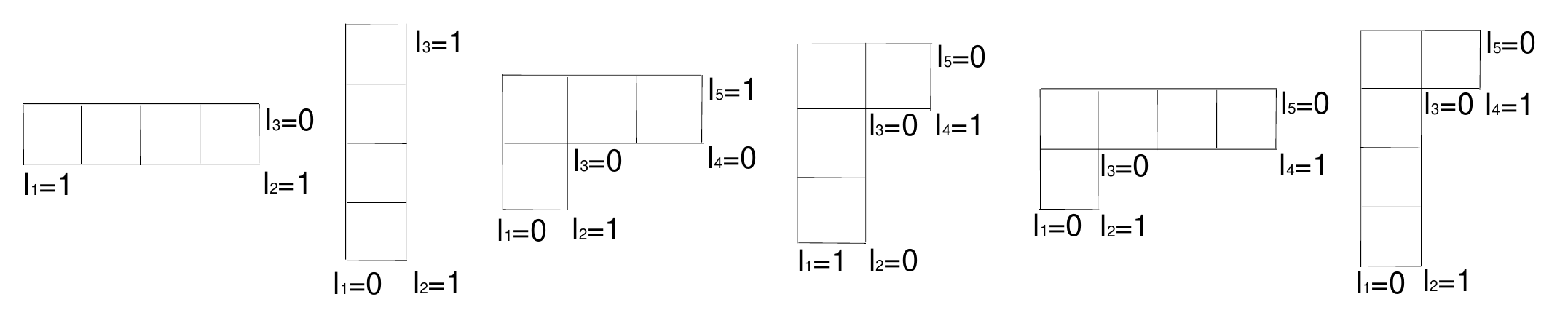}
\caption{Gravity solutions for $k\!=\!2$, $N\!=\!2$ containing fractional anti
M2-branes: all torsions are brought to $\ell\!=\!0,1$.}\label{k2n2negative}
  \end{center}
\end{figure}
There are more gravity solutions with fractional anti M2-branes at the convex corners.
These solutions at $N\!=\!2$ are shown in Fig \ref{k2n2negative}. The torsion $\ell$ ranged
in $0\!\leq\!\ell\!<\!k$ at a convex corner carries the M2-charge $-\frac{\ell(k\!-\!\ell)}{2k}$
for anti M2-branes. For instance, the last Young diagram in Fig \ref{k2n2negative}
has the M2-charge
\begin{equation}
  N=\frac{\#({\rm boxes})}{k}-\frac{\ell_2(k\!-\!\ell_2)}{2k}-\frac{\ell_4(k\!-\!\ell_4)}{2k}
  =\frac{5}{2}-\frac{1}{4}-\frac{1}{4}=2\ .
\end{equation}
Similar calculations yield $N\!=\!2$ for all other cases.

\begin{figure}[t!]
  \begin{center}
    \includegraphics[width=10cm]{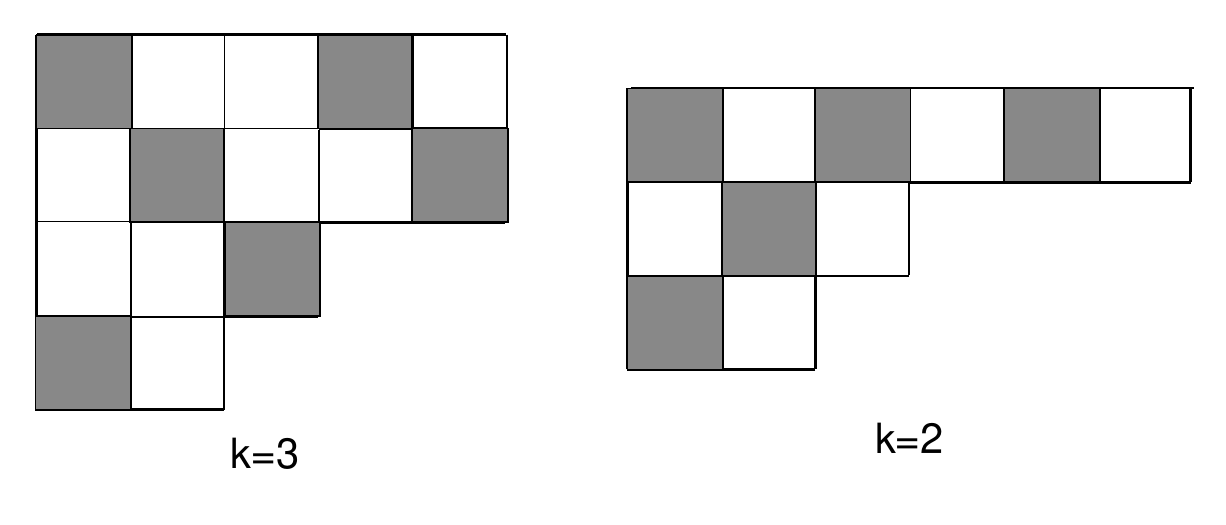}
\caption{Examples of blended Young diagrams for $k\!=\!3$ (left) with $N\!=\!6$ and
$k\!=\!2$ (right) with $N\!=\!5$. The M2-brane number $N$ is given by the number of
grey boxes.}\label{blended}
  \end{center}
\end{figure}

For all gravity solutions illustrated in Figs \ref{k2n1}, \ref{k2n2}, \ref{k2n2negative} by
Young diagrams, and also for many other examples that we have checked with higher $k$ and $N$,
we empirically find that the M2-brane number $N$ is always given by the
size of the so-called \textit{blended Young diagram} with charge $k$
(the Chern-Simons level). The size of a $k$-blended Young diagram is the number
of boxes in a diagram forming diagonal lines with separation $k$, as shown in Fig
\ref{blended}. For instance, for the left Young diagram at $k\!=\!3$, we find
$(\ell_1,\ell_2,\ell_3,\ell_3,\ell_4,\ell_5,\ell_6,\ell_7)\!=\!(2,2,2,0,1,0,2)$ and
\begin{equation}
  N\!=\!\frac{15}{3}+\frac{1}{3}-\frac{1}{3}+\frac{1}{3}-0+\frac{1}{3}-0+\frac{1}{3}
  \!=\!6\ ,
\end{equation}
which equals the size of the 3-blended partition (number of grey boxes). See, for
instance, \cite{Nekrasov:2003rj,Dijkgraaf:2007fe} for more explanations on the blended
Young diagrams.
\begin{figure}[t!]
  \begin{center}
    \includegraphics[width=10cm]{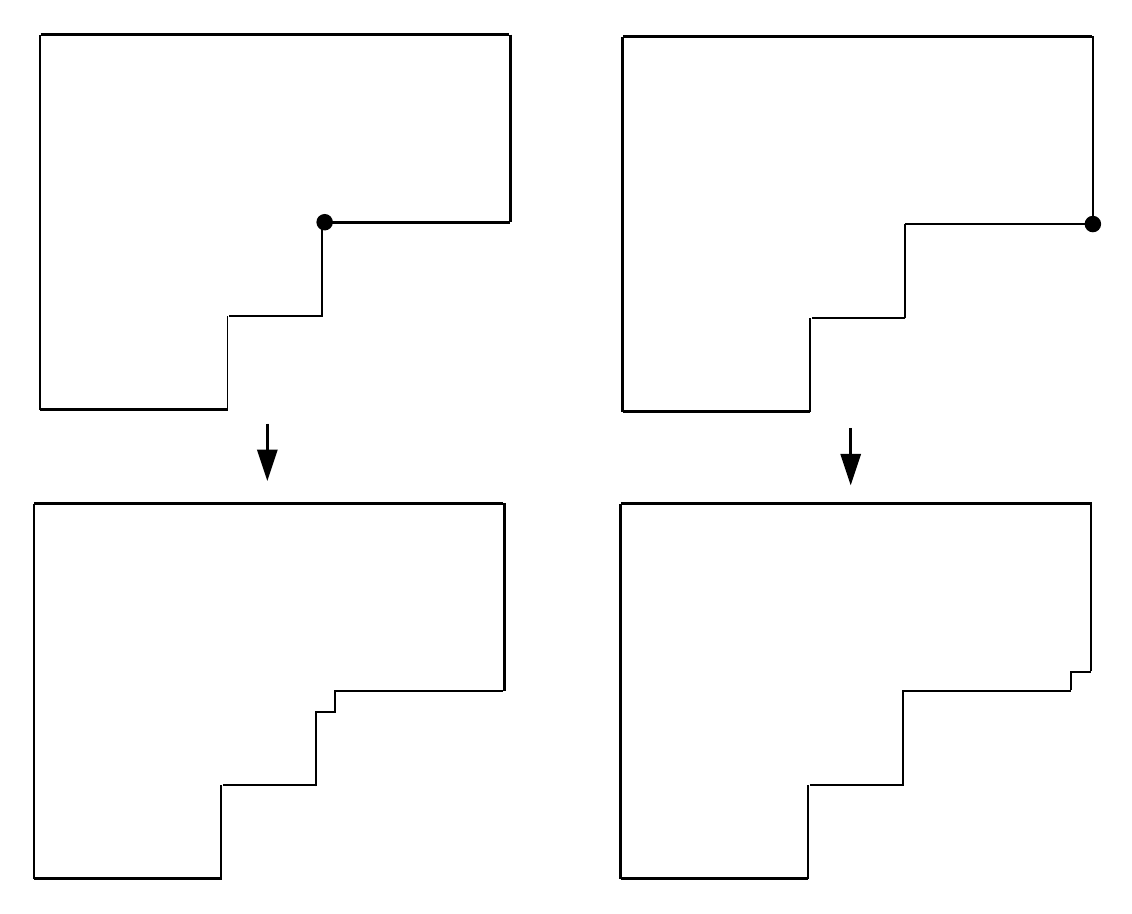}
\caption{Probe M2-branes and anti M2-branes (dots on the left/right sides) can be
geometrized by putting or eliminating a box at the concave/convex corner of the Young
diagram.}\label{redundant}
  \end{center}
\end{figure}

Now let us consider the meaning of the gravity solutions which contain negatively
charged fractional M2-branes. It is illustrative to start from a simpler case at
$k\!=\!1$, and put some full probe M2- and/or anti M2-branes in the background of
\cite{Lin:2004nb}. An M2-brane in a background given by a droplet (Young diagram) is
stabilized at one of the concave corners of the Young diagram, as the latter correspond
to the odd edges of the droplet.
As the full set of $\mathcal{N}\!=\!8$ ground states is given by the solution of
\cite{Lin:2004nb} parametrized by Young diagrams, adding an extra probe M2-brane at
a concave corner provides a redundant description of the ground states. One can naturally
identify the corresponding gravity solution which fully geometrizes the probe M2-brane.
In the example shown on the left side of Fig \ref{redundant}, one geometrizes the probe
by removing it and attaching an extra box at the same concave corner. Similarly, a
BPS anti M2-brane with negative M2-brane charge at a convex corner can be geometrized
by eliminating one box at that corner. Obviously, both configurations with probe
M2- or anti M2-branes are redundant.

At general $k\neq 1$, fractional anti M2-branes cannot be fully geometrized,
as we have been discussing in section 2. Positively charged fractional M2's are
indeed very essential for obtaining the gravity solutions dual to the field theory
vacua. However, we find that all negatively charged fractional M2's at convex
corners can be eliminated to yield a geometry (Young diagram) with less
boxes plus some extra fractional M2's at concave corners. For instance, at
$k\!=\!2$, one can eliminate a box at each convex corner hosting nonzero torsion
from the six Young diagrams in Fig \ref{k2n2negative}, to obtain one of the Young
diagrams in Fig \ref{k2n2}. They reduce to
\begin{equation}
  \Yboxdim7pt\yng(4)\ ,\ \yng(1,1,1,1)\ ,\ \yng(3,1)\ ,\ \yng(2,1,1)\ ,\ \yng(4,1)
  \ ,\ \yng(2,1,1,1)\ \rightarrow\ \ \yng(3)\ ,\ \yng(1,1,1)\ ,\ \yng(3)\ ,\
  \yng(1,1,1)\ ,\ \yng(3)\ ,\ \yng(1,1,1)\ .
\end{equation}
For $k\geq 3$, one sometimes has to eliminate more than one box at a convex
corner. For instance, at $k\!=\!3$, we find the reduction
\begin{equation}
  \Yboxdim10pt\young(\bullet\hfil\hfil\bullet\hfil\hfil,\hfil\bullet\hfil\hfil
  \bullet,\hfil\hfil\bullet\hfil,\bullet\hfil,\hfil,\hfil)\ \rightarrow\ \
  \young(\bullet\hfil\hfil\bullet\hfil,\hfil\bullet\hfil\hfil\bullet,\hfil
  \hfil\bullet,\bullet)\ .
\end{equation}
The number of black dots in a diagram is the M2-brane number $N$
($\!=\!6$ in this case) from the blended Young diagram for $k\!=\!3$. On the
left/right side, the discrete torsions are
\begin{equation}
  (\ell_1,\ell_2,\cdots,\ell_{11})\!=\!(0,{\bf 2},0,{\bf 2},2,{\bf 2},0,{\bf 2},2)
  \ ,\ \ (\ell_1,\ell_2,\cdots,\ell_7)\!=\!(2,{\bf 0},1,{\bf 0},1,{\bf 0},2)\ ,
\end{equation}
where the bold-faced numbers are for anti M2-branes.
The general rule that we find for eliminating the boxes from
convex corners is to remove the maximal number of boxes without
removing any box which contains a black dot.

\section{Charged particles and their gravity duals}

Having studied the vacua of this theory, it is then of interest to study the
elementary excitations. In particular, we would first like to study the massive
particles in various vacua, charged under the unbroken gauge group (\ref{unbroken}).
When all 't Hooft coupling constants $\frac{N_n}{k}$, $\frac{N_n^\prime}{k}$
are small, one can simply diagonalize the mass matrix to understand their
spectrum classically. This is subject to quantum correction when some couplings
are large. For the BPS particles in the supersymmetric vacua
with $N_n,N_n^\prime\leq k$, one may expect to understand or test some aspects
of the gauge/gravity duals that we proposed in the previous sections without much
difficulty, hopefully relying on the weakly coupled field theory results.

However, even the BPS spectrum should also be studied with care. The first
signal for the subtlety is that the Chern-Simons theory with $U(k)_{\pm k}$ gauge group
and level (with $\mathcal{N}\!=\!3$ supersymmetric UV completion given by Yang-Mills
Chern-Simons theory) is dual to nothing \cite{Aharony:2008gk} at low energy. This
happens when the 't Hooft coupling becomes $1$. Therefore, naive classical spectrum
for charged particles at small 't Hooft coupling should break down
even for BPS modes, whenever the relevant gauge group becomes $U(k)$. It is
possible that the BPS spectrum could be nontrivial for particles charged under
$U(\ell)_k$ when $\ell\sim k$. On the other hand, in some strongly coupled regime,
the Seiberg like duality could allow one to study the spectra in dual descriptions
with small coupling constants.

Turning to the charged particles from the gravity dual, it is
natural to seek for such states from open M2-branes having two ends on fractional
M2-branes in the background. In particular, we would like to have them wrap the
`M-theory circle' which is subject to the $\mathbb{Z}_k$ orbifold. By doing so,
the open M2-branes reduce to fundamental strings heading towards two
$\mathbb{R}^8/\mathbb{Z}_k$ singularities at large $k$. Near the tips of the orbifolds,
the world-volumes of open M2's ending on fractional M2-branes will be locally same as
the open membranes dual to W-bosons in the Coulomb phase of conformal Chern-Simons-matter
theories, studied in \cite{Berenstein:2008dc}.

In section 4.1, we study classical configurations for the probe open membranes
preserving half of the supersymmetry (6 real). Classically, their masses are
given by the area of the membranes. There could
in principle be subtle corrections from the two end points, over which we have
little control. The classical analysis would be reliable when the membranes are
macroscopic. In this case, one finds a good agreement with the results from
weakly coupled field theory. See our section 4.1 for the subtle subleading terms
and some comments on them.

In section 4.2, we briefly discuss possible non-relativistic conformal
field theories in this gauge/gravity duality, as they have to do with taking the low
energy limit keeping a subset of these massive particles. At each vacuum,
one can take the non-relativistic limit keeping either `particle' or `anti-particle'
modes with respect to a $U(1)$ symmetry (to be identified with the particle
number symmetry). The geometric realizations \cite{Son:2008ye,Balasubramanian:2010uw}
of holographic non-relativistic conformal symmetry do not appear to be relevant in our
context, implying broader possibilities of holographic non-relativistic systems than
those studied so far.

\subsection{Spectrum of charged particles}

Before presenting the analysis for the BPS spectrum of charged particles, we
note that the vacua preserve global $SU(2)_1\times SU(2)_2$ R-symmetry by mixing
them with some global parts of
the $U(N)\times U(N)$ gauge symmetry. For the $m\times (m\!+\!1)$ matrix block
of first type, $SU(2)_1$ is realized on the $m$ and $m\!+\!1$ rows and columns of
the gauge indices as totally symmetric representations. Similarly,
for the $(m\!+\!1)\times m$ matrix block of second type, $SU(2)_2$ is realized on
the $m\!+\!1$ and $m$ rows and columns as totally symmetric representations.
Detail on the tensor representations of the irreducible blocks is
explained in appendix \ref{tensor}.

We first study the massive vectors modes, which acquire nonzero masses
through Higgs mechanism. We start by considering a classical vacuum containing
an irreducible block of first type which is an $m\times(m\!+\!1)$ matrix, and
another block of first type which is an $n\times(n\!+\!1)$ matrix. We find that
the `off-diagonal' $m\times n$ matrix in $A_\mu$ and the $(m\!+\!1)\times(n\!+\!1)$
matrix in $\tilde{A}_\mu$ mix with each other in the Gauss' law, which becomes
the classical equation of motion for massive vectors.
The modes solving the Gauss' law equation, after an appropriate
gauge-fixing for scalars to be eaten up by vectors, have following masses and
representations under $SU(2)_1$:
\begin{equation}
  j_1\!=\!\frac{m\!+\!n\!-\!2p}{2}:\ \ M=\frac{2\pi\mu}{k}(m\!+\!n\!-\!2p)
  \ \ {\rm or}\ \ \frac{2\pi\mu}{k}(m\!+\!n\!+\!2\!-\!2p)\ ,
\end{equation}
where half-integral $j_1$ denotes the `total angular momentum' quantum number for $SU(2)_1$,
and an integer $p$ runs over $1\!\leq\!p\!\leq\min(m,n)$. Namely, there are two
representations with given $j_1$ quantum number with different masses.
Exceptionally, there is a representation with $p\!=\!0$ (i.e. $j_1\!=\!\frac{m\!+\!1}{2}$)
with the mass $\frac{2\pi\mu(m\!+\!n)}{k}$. See appendix \ref{tensor} for the derivation.
One can do a similar analysis for the complex conjugate modes, after which we obtain
the same representations and masses: this is obvious from the complex conjugation
of the results above. After quantization, with the Chern-Simons term providing
the symplectic 2-form, one of the conjugate pair would become the creation operator while
the other becomes the annihilation operator.

The above result contains both BPS and non-BPS modes. The modes preserving some
supersymmetry can be analyzed by studying the fermion supersymmetry variations. In the
convention of appendix \ref{mass}, the $\mathcal{N}\!=\!6$ supersymmetry is labeled by spinors
$\xi^{\alpha\beta}$ with $\alpha,\beta\!=\!1,2,3,4$ in the anti-symmetric representation
of $SU(4)$, broken to $SU(2)_1\times SU(2)_2\times U(1)\times U(1)_b$ by the mass deformation.
The spinor is subject to the reality condition
$(\xi^{\alpha\beta})^\ast=\frac{1}{2}\epsilon_{\alpha\beta\gamma\delta}\xi^{\gamma\delta}$.
We investigate the $\frac{1}{6}$-BPS modes preserving $\xi^{12}_-$ and its conjugate
$\xi^{34}_+$, where the subscript $\pm$ denotes the spatial spin part which
diagonalizes $\gamma^0\xi_\pm^{\alpha\beta}=\mp i\xi_{\pm}^{\alpha\beta}$.
As analyzed in appendix \ref{susy-gravity}, these $\frac{1}{6}$-BPS
particles correspond to open M2-branes (or fundamental strings in the type IIA limit)
which are smeared along $S^2\times\tilde{S}^2$ base of $S^3\times\tilde{S}^3$. The open
M2-branes localized on these 2-spheres are shown to be $\frac{1}{2}$-BPS from the
gravity dual, which is broken to $\frac{1}{6}$ after this smearing.\footnote{This is
very similar to $\frac{1}{6}$-BPS Wilson loops in $AdS_4\times S^7/\mathbb{Z}_k$ smeared
on $\mathbb{CP}^1\subset\mathbb{CP}^3$. See appendix \ref{susy-gravity}.} As we shall
see in appendix \ref{mass}, our charged particles carry definite $SU(2)_1\times SU(2)_2$
charges, inevitably delocalized on the 2-spheres. The modes preserving this supersymmetry
exist only when $p\!=\!\min(m,n)$, and among the above two representations, only one with
\begin{equation}
  M_{BPS}=\frac{2\pi\mu|m-n|}{k}\ \ \ (m\neq n)
\end{equation}
is BPS. This shall be reproduced from gravity when the classical membrane
analysis is reliable.

The analysis of massive vector modes connecting two irreducible blocks of second type,
with sizes $m$ and $n$, is completely analogous to the above modes connecting two
first type blocks of vacuum. The BPS mass is again given by $M_{BPS}\!=\!\frac{2\pi\mu|m\!-\!n|}{k}$
for $m\neq n$.

The massive vectors connecting an $m\times(m\!+\!1)$ irreducible block of first type and
an $(n\!+\!1)\times n$ block of the second type come from the $m\times(n\!+\!1)$ matrix
part of $A_\mu$, and also from $(m\!+\!1)\times n$ matrix of $\tilde{A}_\mu$. The numbers of
rows and columns of the latter two matrices are the dimensions of irreducible representations
of $SU(2)_1$ and $SU(2)_2$, respectively. All masses are given by
\begin{equation}
  M=\frac{2\pi\mu(m\!+\!n\!+\!1)}{k}\ .
\end{equation}
Similar analysis can be done for the complex conjugate $(n\!+\!1)\times m$ block of
$A_\mu$ and $n\times(m\!+\!1)$ block of $\tilde{A}_\mu$, where we (obviously) obtain the
same mass. Combining the conjugate modes and after quantization, one finds
that only half of the two possible modes $A_1\pm iA_2$ remain BPS. See again
appendix \ref{mass} for the details.

Similar analysis for the BPS scalar modes can be done. The modes connecting two
first type blocks, or two second type blocks, of sizes $m$ and $n$ have mass
\begin{equation}
  M_{BPS}=\frac{2\pi\mu|m-n|}{k}\ \ \ (m\neq n)\ ,
\end{equation}
while the modes connecting one first type and one second type blocks of sizes $m$, $n$
have mass
\begin{equation}
  M_{BPS}=\frac{2\pi\mu(m\!+\!n\!+\!1)}{k}\ .
\end{equation}
The BPS masses are the same as the corresponding vector modes. See appendix \ref{mass}.

Now we study the same charged objects from the gravity dual. From the identification of the
gauge symmetry, the natural candidates are the open M2-branes connecting fractional M2-branes
at the orbifold fixed points. Such states exist only when $k\neq 1$.
The open M2's should be extended along a line in the $xy$ plane and wrap the M-theory circle
(i.e. diagonal direction of the two Hopf fibers of two 3-spheres). The resulting
2-dimensional spatial manifold is topologically $S^2/\mathbb{Z}_k$. From the
Nambu-Goto action, the energy of this open M2-brane is
\begin{equation}
  \tau_{M2}\int_0^{\frac{2\pi}{k}}d\phi\int\sqrt{dx^2\!+\!dy^2}
  e^{2\Phi/3}\cdot e^{-\Phi/3}h\cdot \left(2y\cosh G\right)^{1/2}e^{-\Phi/3}
  =\frac{2\pi\tau_{M2}}{k}\int\sqrt{dx^2+dy^2}\ .
\end{equation}
The last line integral is the `length' of the curve in the $xy$ plane with the
Euclidean measure. As the two ends of the curve are at $y=0$ where fractional M2-branes
are located, the energy is minimized when the curve is a straight line $y=0$.
To compare this result with the field theory, let us Seiberg-dualize all the gauge
groups below the Fermi level and label the fractional branes above/below the Fermi
level by $n=0,1,2,\cdots$ as explained in section 2.3, to go to the
same duality frame as the field theory. For the open M2 connecting $m$'th
and $n$'th fractional M2-branes above the Fermi level (with $m\!>\!n$), the mass is given by
\begin{equation}\label{gravity-mass}
  M=\frac{2\pi\tau_{M2}}{k}\frac{2\mu_0}{\pi\tau_{M2}}\left(km\!+\!N_m-kn\!-\!N_n\right)
  =4\mu_0\left(m\!-\!n+\frac{N_m}{k}-\frac{N_n}{k}\right)\ .
\end{equation}
Upon identifying $\mu_0=\frac{\pi\mu}{2k}$, the first two terms in the
parenthesis agrees with the weakly coupled field theory spectrum. The last two terms
are 1-loop corrections. We are not completely sure which of the two calculations
are valid between classical field theory and gravity at this subleading level. Strictly
speaking, both calculations
are valid when $N_m,N_n\ll k$: otherwise, as there are sectors of strongly coupled
Chern-Simons theory both from field theory and gravity sides, both calculations might be
unreliable. Also, for classical M2-brane calculation to be reliable, we should have
$m-n\gg 1$ for the worldvolume of open M2's to be macroscopic. When all the requirements
for the validity of gauge/gravity calculations are satisfied, we find that the two results
agree with each other.

At this point, we should mention that there are examples in which BPS masses receive
quantum corrections at 1-loop level. In particular, in the context of Chern-Simons-matter
theories, such a phenomenon is observed for BPS vortices when a $U(1)$ symmetry is broken,
together with the same correction to the central charge.\footnote{
We thank Choonkyu Lee for informing us of this result, which is in his unpublished
work.} It is not clear whether this implies that subleading terms in (\ref{gravity-mass}) should
be seriously considered. More comments about the exact BPS spectrum is given at the end of this
subsection, and also in the discussion section.

Similar analysis for the charged particles can be done for the open M2's
connecting fractional M2-branes below the Fermi level. The analysis is completely
analogous to the above.

The last case to discuss is the modes connecting
$m$'th fractional M2-brane above the Fermi level, and $n$'th fractional
M2-brane below the Fermi level. Repeating the analysis above, one
obtains the following mass from the classical M2-branes:
\begin{equation}
  M=4\mu_0\left(m\!+n\!+\frac{N_m}{k}\!+\!\frac{N_n^\prime}{k}\right)\ .
\end{equation}
Comparing with the field theory spectrum $\propto (m\!+\!n\!+\!1)$, we first find a
discrepancy proportional to the 't Hooft couplings, which can be ignored in the weak
coupling limit. Even after ignoring this part, another difference is that the field theory
spectrum has $+1$ compared to the classical gravity result. This is again subleading in the
limit $m\!+\!n\gg 1$ in which classical membrane analysis is reliable, again implying
agreement between the two calculations.

Again it could be that the last subleading $+1$ discrepancy between the field theory and
gravity may have to be considered seriously, and one may want to clarify which of the two
calculations has to be refined. As the field theory calculation is obviously reliable when
't Hooft couplings are small, we suspect that classical gravity calculation has to be corrected,
perhaps by taking into account some effects from the boundary of open M2-branes. A similar
subleading discrepancy was observed in the conformal $\mathcal{N}\!=\!6$
Chern-Simons-matter theory, in the study of the field theory dual of giant M2-brane torus.
The last object can be regarded as open strings connecting spherical D2-brane dual giant
gravitons \cite{Berenstein:2009sa}, somewhat similar to our open M2's as they are
connecting spherically polarized M5-branes in the probe limit \cite{Bena:2000zb}.

As shown in appendix \ref{susy-gravity}, the open M2-branes localized on $S^2\times\tilde{S}^2$
base of $S^3\times\tilde{S}^3$ are $\frac{1}{2}$-BPS. On the other hand, as we smear the
open M2-branes along the transverse 2-sphere, one only obtains $\frac{1}{6}$-BPS solutions.
The supersymmetry of the gravity and field theory sides thus agree with each other.

We remark that we have not clarified the degeneracies of various BPS particles from
the gravity dual. One point which could be important is that open membranes are transverse
to one of the two 2-spheres spanned by either $SU(2)_1$ or $SU(2)_2$ R-symmetries. By
examining the 4-form flux of (\ref{llm}), we find that open M2's effectively behave as
charged particles moving on $S^2$ in nonzero magnetic fields. There are nontrivial
degeneracies from the lowest magnetic monopole spherical harmonics. It could be possible to
understand the degeneracies from the gravity dual in this way, which we leave as a future work.

Another point worth a consideration is that the gravity solution does not see the
`Fermi level' at all. Before performing Seiberg-like dualities for some fractional M2-branes
in section 2.3, all fractional M2-branes are described by $U(\ell)_{-k}$
type Chern-Simons theories, where $\ell$ is the torsion at each orbifold fixed point.
We have performed Seiberg-like dualities for the Chern-Simons theories living below the Fermi level
of the droplet, to compare the gravity result with the field theory. The field theory in the
last Seiberg-duality frame is weakly coupled when all $N_n$, $N_n^\prime$ are much smaller than $k$.
In the gravity viewpoint, this means that the quantized lengths of all black regions above the
Fermi level are much smaller than $k$, while the lengths of all white regions below the Fermi level are
smaller than $k$.\footnote{Of course, such gravity solutions have Planck scale curvatures.}
So the notion of Fermi level emerges just by demanding the field theory be weakly coupled
in a particular duality frame. At the full non-perturbative level, we think the field
theory should also be ignorant about the Fermi level. To test this, one could
rely on localization methods to calculate the partition function (or index) of all BPS particles.
Similar to the instanton partition functions \cite{Nekrasov:2003rj} in 5 dimensional
Yang-Mills theories compactified on a circle, we may introduce an Omega deformation with the
$SO(2)$ rotation on spatial part of $\mathbb{R}^{2,1}$ to lift the translational zero modes
of these particles. This partition function would include contributions from various
vortices \cite{Kim:2009ny,Auzzi:2009es} as well as elementary particles. Such a calculation
could also resolve the puzzle raised in \cite{Auzzi:2009es} about the zero modes (or
more precisely ground state degeneracies) of vortices.

\subsection{Remarks on non-relativistic conformal symmetry}

The massive charged particles studied in the previous section are subject to
various conservations of $U(1)$ charges. In particular, the overall $U(1)$ factors
in the $U(N_n)$ or $U(N_n^\prime)$ type gauge groups turn out to be important
when $N_n,N_n^\prime\neq 0,k$. As various massive particles are bi-fundamental
in two factors of such unitary groups, the charge of a particle under a given
$U(1)$ is either $+1$, $0$ or $-1$.

In the symmetric vacuum, with $N_0\!=\!N_0^\prime\!=\!N$ in our notation,
a non-relativistic limit was considered in \cite{Nakayama:2009cz,Lee:2009mm} which
discards `anti-particle modes' which are negatively charged under the baryon-like
$U(1)_b$. See also \cite{Nakayama:2009ed} for further
discussions on this theory. As $U(1)_b$ is the global part of the difference
between two overall $U(1)$'s in $U(N)\times U(N)$ \cite{Aharony:2008ug},
$U(1)_b$ is a special combination of many $U(1)$'s in the general Higgs vacua
mentioned in the previous paragraph.
Similar non-relativistic limits can be taken in the Higgs vacua
that we have been discussing, in which scalar expectation values are nonzero.
With various overall $U(1)$ factors, one can keep the modes with definite signs
of charges under various $U(1)$'s and take a non-relativistic limit
similar to \cite{Nakayama:2009cz,Lee:2009mm}. In the bosonic sector, one would
obtain non-relativistic kinetic terms and quartic interaction
terms of various modes. This system would have a scale symmetry
with dynamical exponent $z\!=\!2$. It should also have Galilean boost symmetry
as well as the symmetries analogous to particle number $U(1)$ mentioned
above. In \cite{Nakayama:2009cz,Lee:2009mm}, the non-relativistic system also
has time special conformal symmetry. This appears when the relative coefficients of
the Chern-Simons term, kinetic term and the quartic interactions are fine tuned.
The general non-relativistic system from the Higgs vacua could also have this symmetry,
although we have not performed this analysis.

The particle number-like $U(1)$ symmetries are realized in the bulk as part of the
gauge symmetry localized on fractional M2-branes. Note that this is in contrast to
\cite{Son:2008ye}, in which an isometry of a spacetime
direction is used to realize the particle number symmetry. This is because our
gravity solutions for mass-deformed M2-branes do not fully geometrize the
M2-brane charges into flux, but leave some of them as fractional M2-branes.
We expect that other part of non-relativistic symmetries will not be geometrically
realized, either. For instance, in \cite{Son:2008ye}, other symmetries
also have nontrivial actions along the direction which realizes particle number
isometry. Our system is also different from another recently studied
non-geometric realization of the particle number \cite{Balasubramanian:2010uw},
as our particle symmetry lives in $2\!+\!1$ dimensions.

After the suggestion of \cite{Son:2008ye} on the gravity duals
of non-relativistic conformal systems, there have been some attempts to obtain
such solutions arising from mass-deformed M2-branes \cite{Ooguri:2009cv},
having the symmetric vacuum theory of \cite{Nakayama:2009cz,Lee:2009mm} in mind.
These works sought for solutions with some
supersymmetry, and also with the $SU(2)_1\times SU(2)_2$ isometry as a gravity dual
of the R-symmetry of the $\mathcal{N}\!=\!6$ system. Firstly, if one is studying
the symmetric vacuum, one would have to study solutions with broken supersymmetry,
as the field theory vacuum breaks supersymmetry for $N>k$.
Also, the reason why $SU(2)_1\times SU(2)_2$ symmetry is imposed in
\cite{Ooguri:2009cv} is because $SO(4)\times SO(4)$ R-symmetry of the field
theory at $k\!=\!1$ is reduced to $SU(2)\times SU(2)$ for general $k$. If this reduction
of symmetry happens by a $\mathbb{Z}_k$ orbifold from the gravity dual like the
solutions in this paper, then the ansatz one should consider
is much more restrictive than those considered in \cite{Ooguri:2009cv}.
It would be interesting to see if such a strong restriction would help us
find non-supersymmetric gravity solutions.

\section{Discussions}

In this paper, we have shown that $\mathbb{Z}_k$ quotients of the polarized M2-brane
solutions of \cite{Lin:2004nb} are dual to the supersymmetric vacua of the mass-deformed
$\mathcal{N}\!=\!6$ Chern-Simons-matter theories, after taking fractional
M2-branes and strong coupling Chern-Simons dynamics into account.

As the gravity duals that we found in this paper have rich structures, we think this model
can be studied for various purposes. In particular, it would be very interesting to
see if we can extend or refine the studies on quantum Hall systems based on conformal
Chern-Simons-matter theories \cite{Fujita:2009kw}, to the theories with
mass gap. The latter should be desirable as the quantum Hall systems are
gapped in the bulk. Also, it would be interesting to study various interface or defect
configurations which support stable massless charged states at edges.
$D4$ and $D8$ branes are used in CFT-based models \cite{Fujita:2009kw}. As the mass-deformed
geometry also has topological 4-cycles, M5-branes wrapping them would naturally provide
$1\!+\!1$ dimensional defects in $\mathbb{R}^{2,1}$. Due to the presence of nonzero magnetic
flux on such cycles, one has to attach M2-branes to the defect for tadpole cancelation
of the 3-form world-volume field. These tadpole M2-branes are extended along half of
$\mathbb{R}^{2,1}$, having the defect as their boundaries.

As explained in this paper with elementary excitations, studies of BPS particles
in this theory expose rich structures and various subtleties. Although the leading energies
for macroscopic open membranes agree with elementary excitation masses, we think it could be important to
understand the exact spectrum from both sides. Another type of subtlety in the BPS spectrum
was found in \cite{Auzzi:2009es} in the studies of vortex solitons and D0-branes in the
gravity dual. Namely, it has been shown that the dimensions of moduli spaces of the dual objects
apparently seem to disagree, where the field theory and gravity show complex $1$ and
$3$ dimensional moduli, respectively. This difference would yield different ground state
degeneracies after quantization. To get an exact result for the spectrum of BPS
particles in the field theory side, one can try a localization calculation for the index
for these BPS states, similar to the instanton partition function counting BPS bound states
of 5 dimensional supersymmetric Yang-Mills theories \cite{Nekrasov:2003rj}.

As discussed in section 3.2, it may not be too difficult to seek for the
gravity solutions with non-relativistic conformal symmetry dual to the symmetric
vacuum, breaking supersymmetry.

For $\mathcal{N}\geq 4$ supersymmetric mass-deformed Chern-Simons-matter theories,
the S-matrices for the scattering of 2 to 2 particles in the 't Hooft limit have
been studied in \cite{Agarwal:2008pu} for the symmetric vacuum. In the Higgs phase
vacua, various macroscopic open membranes could be used to carry out some semi-classical
calculations. It would be interesting to study them, firstly in the $\mathcal{N}\!=\!6$
theories discussed in this paper. Also, as mentioned in \cite{Agarwal:2008pu},
S-matrix of the mass-deformed theory is well defined, in contrast to the conformal theories.

It is also possible to replace the $\mathbb{Z}_k$ orbifold by other orbifolds to
yield the gravity duals of mass-deformed Chern-Simons-matter theories with reduced
supersymmetry. The conformal $\mathcal{N}\!=\!5$ or $4$ supersymmetric theories have moduli
spaces given by orbifolds of $\mathbb{R}^8$. The mass-deformed geometries could simply be
obtained by acting the same orbifold quotients on the $\mathcal{N}\!=\!8$ solution of
\cite{Lin:2004nb}. For instance, the $\mathcal{N}\!=\!5$ theory replaces $\mathbb{Z}_k$ by
a dihedral group \cite{Hosomichi:2008jb,Aharony:2008gk}. To classify and study these vacua
from the gravity duals like our $\mathcal{N}\!=\!6$ vacua, one should know the contributions
of orbifold curvature singularities and the fractional M2-branes to the M2-brane charge. This
would be a generalization of \cite{Bergman:2009zh,Aharony:2009fc} to other orbifolds, which
is not done yet. The classical vacua of the $\mathcal{N}\!=\!5,4$ mass-deformed field theories
are not fully classified either, as far as we are aware of.

\vskip 0.5cm

\hspace*{-0.8cm} {\bf\large Acknowledgements}

\vskip 0.2cm

\hspace*{-0.75cm} We would like to thank Dongmin Gang, O-Kab Kwon, Choonkyu Lee,
Sangmin Lee, Sungjay Lee, Eoin O Colgain, Soo-Jong Rey, Tadashi Takayanagi and
especially Ki-Myeong Lee for helpful discussions. We also thank Sungjay Lee for comments
on the preliminary version of this manuscript. This work is supported in part by
the Research Settlement Fund for the new faculty of Seoul National University (SK),
the BK21 program of the Ministry of Education, Science and Technology (HK, SK),
the National Research Foundation of Korea (NRF) Grants No. 2010-0007512 (SC, HK, SK),
2007-331-C00073, 2009-0072755 and 2009-0084601 (HK).

\appendix

\section{Supersymmetry of gravity solutions and M2-branes}\label{susy-gravity}

In this appendix, we explain the $\mathcal{N}\!=\!8$ Killing spinor of (\ref{llm}),
the $\mathcal{N}\!=\!6$ projection after $\mathbb{Z}_k$, the supersymmetry of
probe M2- and anti M2-branes wrapping $\mathbb{R}^{2,1}$, and finally the supersymmetry
of open M2-branes dual to the charged particles.

We take the 11 dimensional vielbein to be
\begin{equation}
  e^\mu\!=\!e^{2\Phi/3}dx^\mu,\ e^3\!=\!e^{-\Phi/3}hdy,\ e^4\!=\!e^{-\Phi/3}hdx,\
  e^{a\!+\!4}\!=\!e^{-\Phi/3}\sqrt{ye^G}\frac{\sigma_R^a}{2},\ e^{a\!+\!8}\!=\!e^{-\Phi/3}
  \sqrt{ye^{-G}}\frac{\tilde\sigma_R^a}{2}\ ,
\end{equation}
where $\mu\!=\!0,1,2$ and $a\!=\!1,2,3$. The right 1-forms $\sigma^a_R$ on $S^3$ are
\begin{equation}\label{right-1-form}
  \sigma^1_R=-\sin\psi d\theta+\cos\psi\sin\theta d\phi\ ,\ \ \sigma^2_R=
  \cos\psi d\theta+\sin\psi\sin\theta d\phi\ ,\ \ \sigma^3_R=\psi+\cos\theta d\phi\ ,
\end{equation}
which satisfy $d\sigma_R^a\!=\!\frac{1}{2}\epsilon^{abc}\sigma^b_R\wedge\sigma^c_R$ and
\begin{equation}
  \frac{i}{2}\tau^a\sigma^a_R=U^\dag dU\ ,\ \ U=\left(\begin{array}{cc}
  \cos\frac{\theta}{2}e^{i\frac{\psi+\phi}{2}}&\sin\frac{\theta}{2}e^{-i\frac{\psi-\phi}{2}}\\
  -\sin\frac{\theta}{2}e^{i\frac{\psi-\phi}{2}}&\cos\theta e^{-i\frac{\psi+\phi}{2}}
  \end{array}\right)\ .
\end{equation}
$\tilde\sigma^a_R$ on $\tilde{S}^3$ are similarly defined.
11 dimensional gamma matrices are taken to be
\begin{equation}
  \Gamma_\mu\!=\!\tau_\mu\otimes{\bf 1}_2\otimes{\bf 1}_2\otimes\gamma_5,\
  \Gamma_{3,4}\!=\!{\bf 1}_2\otimes{\bf 1}_2\otimes{\bf 1}_2\otimes\gamma_{1,2},\
  \Gamma_{a\!+\!4}\!=\!{\bf 1}_2\otimes\sigma_a\otimes{\bf 1}_2\otimes\gamma_3,\
  \Gamma_{a\!+\!7}\!=\!{\bf 1}_2\otimes{\bf 1}_2\otimes\sigma_a\otimes\gamma_4\ .
\end{equation}
where $\gamma_{1,2,3,4}$ and $\tau_\mu$ are $SO(4)$ and $SO(2,1) $ gamma matrices,
respectively.

The $\mathcal{N}\!=\!8$ supersymmetry of the gravity solutions discussed in this paper
is studied in \cite{Bena:2004jw}. We independently checked all equations we use below.
The algebraic half-BPS condition is given by
\begin{equation}\label{half-BPS}
  0\!=\!\frac{1}{2}\left[1+p_1\Gamma^{012}+p_2\Gamma^{01289,10}-p_3\Gamma^{012567}\right]
  \epsilon=\frac{1}{2}\left[1+p_1\gamma_5+ip_2\gamma_{123}+ip_3\gamma_{124}
  \right]\epsilon\ ,
\end{equation}
The coefficients satisfy
\begin{equation}\label{susy-coefficient}
  p_1^2+p_2^2+p_3^2=1\ ,\ \ p_2^2+p_3^2=1-h^{-4}V^2\ .
\end{equation}
The sign of $p_1$ changes as one crosses a curve $p_2^2+p_3^2=1$ on the $xy$ plane.

The Killing spinor can be written as
$\epsilon\!=\!\epsilon^\prime_2\otimes\eta_2\otimes\tilde\eta_2\otimes\psi_4$, following
the gamma matrix decomposition. It turns out \cite{Lin:2004nb,Bena:2004jw} that the $16$
Killing spinors can be decomposed into $8\!+\!8$, depending on the sign of the conditions
satisfied by the two dimensional spinors $\eta$, $\tilde\eta$ on $S^3$, $\tilde{S}^3$:
\begin{equation}\label{S3-spinor}
  \nabla_a\eta_\pm=\pm\frac{i}{2}\sigma_a\eta_\pm\ ,\ \
  \nabla_{\tilde{a}}\tilde\eta_\pm=\pm\frac{i}{2}\sigma_{\tilde{a}}\tilde\eta_\pm\ .
\end{equation}
The projection conditions to these $8$ components are given by
\begin{equation}\label{s3-projection}
  \left[1\pm i\gamma_1\left(\sqrt{ye^{-G}}h\gamma_3+\sqrt{ye^G}h\gamma_4\right)\right]
  \epsilon_\pm\!=\!0\ .
\end{equation}
In the right 1-form basis, the vielbein of a unit 3-sphere is given by
$e^a\!=\!\frac{\sigma^a_R}{2}$. The solutions to (\ref{S3-spinor}) are given by
\begin{equation}
  \eta_+=\eta_{+0}\ ,\ \ \eta_-=U^\dag(\theta,\phi,\psi)\eta_{-0}
\end{equation}
with constant spinors $\eta_{\pm 0}$, and simiarly for $\tilde\eta_\pm$.

The $\mathbb{Z}_k$ quotient shifts $(\psi,\tilde\psi)\rightarrow(\psi\!+\!\frac{4\pi}{k},
\tilde\psi\!+\!\frac{4\pi}{k})$. For the $8$ Killing spinors with $+$ sign,
the components $\eta_+$, $\tilde\eta_+$ are constants. But the frame $e^a$ given by
(\ref{right-1-form}) changes under the shift of $\psi,\tilde\psi$. This implies that the
Killing spinors $\eta_+$, $\tilde\eta_+$ are not invariant under $\mathbb{Z}_k$ shift,
apart from the special cases with $k\!=\!1,2$. On the other hand, the spinors
$\eta_-$, $\tilde\eta_-$ have both components and frame basis changing
under the shift. This structure makes the $-$ spinors all invariant under $\mathbb{Z}_k$.
To see this clearly, we rewrite the $\eta_\pm$ spinors in the left 1-form basis,
$e^a=\frac{\sigma^a_L}{2}$, where
\begin{equation}\label{left-1-form}
  \sigma^1_L=\sin\phi d\theta-\cos\phi\sin\theta d\psi,\ \sigma^2_L=\cos\phi d\theta
  +\sin\phi\sin\theta d\psi,\ \sigma^3_L=d\phi+\cos\theta d\psi\ .
\end{equation}
These satisfy $\frac{i}{2}\tau^a\sigma^a_L=dUU^\dag$. In this basis, it is easy to show
that
\begin{equation}
  \eta_+=U(\theta,\phi,\psi)\eta_{+0}\ ,\ \ \eta_-=\eta_{-0}
\end{equation}
with constant $\eta_{\pm 0}$. Similar expressions can be obtained for $\tilde\eta_\pm$.
As the basis (\ref{left-1-form}) is invariant under the $\mathbb{Z}_k$ shift,
the fact that $\eta_-,\tilde\eta_-$ have constant components imply that the $8$ spinors
with $-$ sign are invariant under $\mathbb{Z}_k$. In this basis, the $\mathbb{Z}_k$ action
on $\eta_+\otimes\tilde\eta_+$ is given by
\begin{equation}
  \eta_+\otimes\tilde\eta_+\rightarrow Ue^{\frac{2\pi i}{k}\sigma_3}\eta_{0+}\otimes
  \tilde{U}e^{\frac{2\pi i}{k}\sigma_3}\tilde\eta_{0+}\ .
\end{equation}
For the spinor to be invariant under this shift, the $\sigma_3$ eigenvalues of
$\eta_{+0}$, $\tilde\eta_{+0}$ should be either $(+,-)$ or $(-,+)$, reducing the $8$
Killing spinors to $4$. Thus, only $12$ out of $16$ spinors
preserve $\mathbb{Z}_k$ in total. The $4$ supercharges with $\eta_+\otimes\tilde\eta_+$
correspond to $Q^{12}$, $Q^{34}$ in the field theory , while the remaining $8$ supercharges
are $Q^{i\tilde{j}}$ with $i\!=\!1,2$, $\tilde{j}\!=\!3,4$ in the notation of appendix \ref{mass}.

\begin{figure}[t!]
  \begin{center}
    \includegraphics[width=10cm]{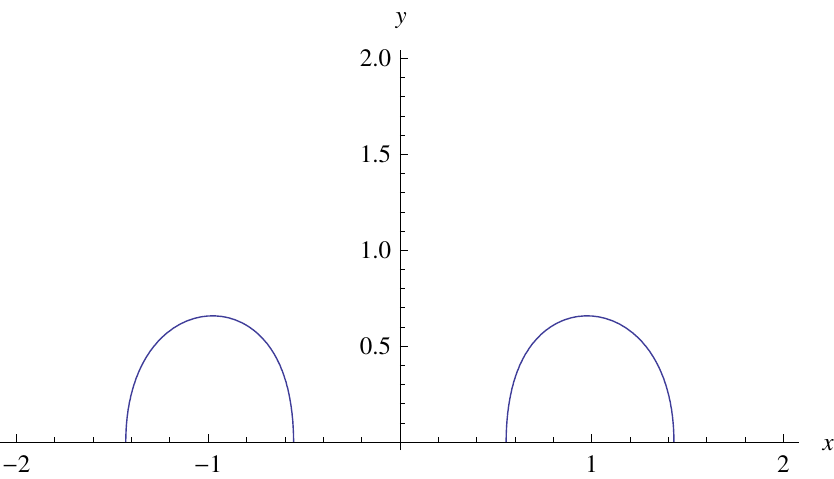}
\caption{The lines on the $xy$ plane with M5-brane supersymmetry, $p_2^2+p_3^2=1$
(or $V(x,y)\!=\!0$), for the droplet $({\bf x}_1,{\bf x}_2, {\bf x}_3,{\bf x}_4,
{\bf x}_5)\!=\!(-2,-1,0,1,2)$. Sign of $p_1$ changes as one crosses the curve.}\label{magnetic-susy}
  \end{center}
\end{figure}

To understand the supersymmetry of full or fractional M2-branes extended along
$\mathbb{R}^{2,1}$, one should know when $\epsilon$ satisfies $\Gamma^{012}\epsilon=\pm\epsilon$,
where $\pm$ sign has to do with whether M2- or anti M2-branes is BPS.
From (\ref{susy-coefficient}), one finds that $p_1^2=1$
when $h^{-4}V^2=1$. This happens at $y\!=\!0$ on the boundaries of black and white
regions: near $y\!=\!0$, $x\!=\!x_i$, one can easily check that
$h,V\sim[(x\!-\!x_i)^2\!+\!y^2]^{-1/2}$, proving our assertion. These are the points
where we argued in section 2.3 that probe M2-branes are located (M2's at $x_{2i\!+\!1}$
and anti M2's at $x_{2i}$). We still have to check the sign of $p_1\!=\!\pm 1$ at
each edge. From the analysis of the potential energy for these branes, we expect that
M2-branes are BPS at odd edges $x_{2i\!+\!1}$ while anti M2-branes are BPS at even edges
$x_{2i}$. This indeed turns out to be the case. To see the change of the $p_1$ sign,
we study the curve $V(x,y)\!=\!0$ on the $xy$ plane, where $p_1\!=\!0$. This curve consists
of various pieces which circles around all even edges $x_{2i}$. See
Fig \ref{magnetic-susy} for an illustration. Firstly, the M2-brane is BPS at asymptotic
infinity. As the edge at $x\!=\!x_1$ is connected to infinity without crossing this curve,
$p_1$ does not change sign and M2-branes are BPS there. Then, moving along the line
$y\!=\!0$ increasing $x$, one finds that $p_1$ at odd and even edges have opposite signs.
This proves our expectation that M2- and anti M2-branes are BPS at odd/even edges, respectively.
We also note that the half-BPS condition (\ref{half-BPS}) locally becomes that for an M5-brane at
$V(x,y)\!=\!0$, as $p_1\!=\!0$ there. It was also shown in \cite{Auzzi:2009es} that probe
D0-branes are stabilized at these curves, which is natural as D0 is mutually supersymmetric
with D4-branes (type IIA reduction of M5).

We finally study the fraction of supersymmetry preserved by the open membranes
discussed in section 4.1. We study those stretched in the white region:
those stretched in the black regions can be analyzed in a similar manner.
The worldvolume supersymmetry demands
\begin{equation}\label{open-m2-susy}
  \epsilon=\Gamma_{047}\epsilon=-\tau_0\otimes\sigma_3\otimes{\bf 1}_2\otimes
  \gamma_{14}\epsilon\ .
\end{equation}
On the white region, the condition (\ref{s3-projection}) for the $\pm$ sector is
$\gamma_{14}\epsilon\!=\!\pm i\epsilon$, respectively. We should thus consider
the projection condition for $\tau_0\otimes\sigma_3$. As the $\tau_0$ projection
on $\mathbb{R}^{2,1}$ spinor is trivial, we should study the $\sigma_3$ projection
of $\eta_\pm$ on $S^3$. Let us first consider the case in which the M2-brane's
position on the $S^2$ base of $S^3$ is localized. Without losing generality, one
can put the M2 at $\theta\!=\!0$. The projection condition (\ref{open-m2-susy})
requires $\eta_\pm$ to take an eigenvalue of $\sigma_3$. In the right 1-form basis,
the $4$ constant spinors containing $\eta_+\otimes\tilde\eta_+$ can be trivially
taken to be such eigenstates, yielding the $\frac{1}{6}$-BPS condition similar to the
field theory charge particles. The $8$ spinors containing $\eta_-\otimes\tilde\eta_-$ at
$\theta\!=\!0$ is given by
\begin{equation}
  \eta_+(\theta\!=\!0,\phi,\psi)\!=\!U^\dag(\theta\!=\!0,\phi,\psi)\eta_{+0}\!=\!
  e^{-i\sigma_3\frac{\psi\!+\!\phi}{2}}\eta_{+0}\ .
\end{equation}
As the $\psi,\phi$ dependent part only contains $\sigma_3$, one can take
$\eta_{+0}$ and $\tilde\eta_{+0}$ to be $\sigma_3$ eigenstates. Therefore,
collecting all sectors with $\pm$ signs, one obtains a half-BPS condition for
open M2-branes localized on $S^2$.

On the other hand, the field theory charged particles discussed in section 4
are all delocalized on $S^2$, as they are in definite angular momentum eigenstates.
For an open M2-brane whose location on $S^2$ base of $S^3$ is smeared,
the $\sigma_3$ projection appearing in (\ref{open-m2-susy}) cannot be imposed at every
point on $S^2$ for $\eta_-\otimes\tilde\eta_-$, as the matrix $U(\theta,\phi,\psi)$
does not commute with $\sigma_3$. So supersymmetry appears only in the $+$ sector,
making this object $\frac{1}{6}$-BPS (with 2 real supercharges from
$\eta_+\otimes\tilde\eta_+$).

The above reduction of $\frac{1}{2}$-BPS to $\frac{1}{6}$-BPS open M2-branes
after smearing is very similar to reduction to the $\frac{1}{6}$-BPS
Wilson loops in $AdS_4\times S^7/\mathbb{Z}_k$ smeared on $\mathbb{CP}^1\subset\mathbb{CP}^3$
\cite{Drukker:2008zx}.

\section{$SU(2)$ tensor representation of the vacua}\label{tensor}

In this appendix, we present the $SU(2)$ tensor representations of the irreducible
vacuum blocks, which is convenient for the group theoretical analysis in appendix \ref{mass}.

Our basic goal is to explain that the $m\times(m\!+\!1)$ irreducible block is
a trivial representation of $SU(2)$, if we regard the $m$ or $m\!+\!1$ indices
for rows/columns as $SU(2)$ symmetric tensors or ranks $m\!-\!1$ and $m$.
To show this, we start by rewriting $m\times(n\!+\!1)$ matrices as
\begin{equation}
  [Z_i]_{a_1a_2\cdots a_{m\!-\!1}}^{b_1b_2\cdots b_{n}}\ ,
\end{equation}
where all $a$ and $b$ type indices are symmetrized and assume either $1$ or $2$.
To get back and forth between the $m\!+\!1$ dimensional vector and rank $m$
symmetric tensor basis with correct normalization, we introduce the normalized
basis $\{|p\rangle\}$ for symmetric tensors
\begin{eqnarray}
  |p\rangle&\equiv&\left(\begin{array}{c}m\\p\end{array}\right)^{-\frac{1}{2}}\left[\frac{}{}\!
  \right.|\stackrel{m\!-\!p}{\overbrace{11\cdots 1}}\ \stackrel{p}{\overbrace{22\cdots 2}}
  \rangle+{\rm relocations}\left.\frac{}{}\!\right]\nonumber\\
  &=&\frac{1}{m!}\left(\begin{array}{c}m\\p\end{array}\right)^{\frac{1}{2}}\left[\frac{}{}\!
  \right.|\stackrel{m\!-\!p}{\overbrace{11\cdots 1}}\ \stackrel{p}{\overbrace{22\cdots 2}}
  \rangle+{\rm all\ permutations}\left.\frac{}{}\!\right]\equiv
  \left(\begin{array}{c}m\\p\end{array}\right)^{\frac{1}{2}}|\tilde{p}\rangle
\end{eqnarray}
for $0\leq p\leq m$, which satisfies $\langle p|q\rangle=\delta_{pq}$.
The symmetrization of a rank $m\!-\!1$ tensor $T$ is attained by the projection
\begin{equation}
  T_{a_1a_2\cdots a_m}\rightarrow
  T^S_{a_1a_2\cdots a_m}=\frac{1}{m!}\delta_{(a_1}^{b_1}\delta_{a_2}^{b_2}\cdots
  \delta_{a_m)}^{b_m}T_{b_1\cdots b_m}=
  T^S_{\stackrel{m\!-\!p}{11\cdots 1}\ \stackrel{p}{2\cdots 2}}=\langle\tilde{p}|T\rangle\ ,
\end{equation}
which is
\begin{eqnarray}
  |T^S\rangle&=&\sum_{a_1\cdots a_m}|a_1\cdots a_m\rangle\langle\tilde{p}|T\rangle=
  \sum_{p=0}^m\left(\begin{array}{c}m\\p\end{array}\right)
  |\tilde{p}\rangle\langle\tilde{p}|T\rangle=
  \sum_{p=0}^m|p\rangle\left(\begin{array}{c}m\\p\end{array}\right)^{\frac{1}{2}}
  T^S_{\stackrel{m\!-\!p}{11\cdots 1}\ \stackrel{p}{2\cdots 2}}\ .
\end{eqnarray}
Here, $m\!-\!p$ and $p$ written above $11\cdots 1$ or $22\cdots 2$ denote
the numbers of repeated indices.
As the normalized basis $\{|p\rangle\}$ for symmetric tensors corresponds
to the normalized basis for the $m\!+\!1$ dimensional unit vectors, the $m\!+\!1$
dimensional vector component $T_p$ (for $p=1,2\cdots,m\!+\!1$) is given in terms of
tensor components as
\begin{equation}
  T_p=\left(\begin{array}{c}m\\p\!-\!1\end{array}\right)^{\frac{1}{2}}
  T^S_{\stackrel{m\!-\!p\!+\!1}{11\cdots 1}\ \stackrel{p\!-\!1}{2\cdots 2}}\ .
\end{equation}
Therefore, the $m\times(n\!+\!1)$ matrix elements $[Z_i]_p^{\ \ q}$ are
given in terms of symmetric tensors by
\begin{equation}
  [Z_i]_p^{\ \ q}=
  \left(\begin{array}{c}m\!-\!1\\p\!-\!1\end{array}\right)^{\frac{1}{2}}
  \left(\begin{array}{c}n\\q\!-\!1\end{array}\right)^{\frac{1}{2}}
  [Z_i]^{\stackrel{n\!-q\!+\!1}{1\cdots 1}\ \stackrel{q\!-\!1}{2\cdots 2}}_{
  \stackrel{m\!-\!p}{1\cdots 1}\ \stackrel{p\!-\!1}{2\cdots 2}}\ .
\end{equation}
In this tensor form, we claim that the $m\times(m\!+\!1)$ vacuum block is given by
\begin{equation}\label{block-tensor}
  [Z_i]_{a_1a_2\cdots a_{m\!-\!1}}^{b_1b_2\cdots b_{m}}=(m\mu)^{\frac{1}{2}}
  \delta_{(i}^{b_1}\delta_{a_1}^{b_2}\cdots\delta_{a_{m\!-\!1})}^{b_m}\ .
\end{equation}
To show this, we study the $m\times(m\!+\!1)$ matrix corresponding to the above
tensor. For $Z_1$, one obtains
\begin{eqnarray}
  [Z_1]_p^{\ \ q}&=&(m\mu)^{\frac{1}{2}}
  \left(\begin{array}{c}m\!-\!1\\p\!-\!1\end{array}\right)^{\frac{1}{2}}
  \left(\begin{array}{c}m\\q\!-\!1\end{array}\right)^{\frac{1}{2}}
  \stackrel{m\!-\!p\!+\!1}{\overbrace{\delta_{(1}^{1}\cdots\delta_{1}^{1}}}\
  \stackrel{p\!-\!1}{\overbrace{\delta_2^2\cdots\delta_{2)}^{2}}}\delta_p^q\nonumber\\
  &=&(m\mu)^{\frac{1}{2}}\left(\begin{array}{c}m\!-\!1\\p\!-\!1\end{array}\right)^{\frac{1}{2}}
  \left(\begin{array}{c}m\\p\!-\!1\end{array}\right)^{-\frac{1}{2}}\delta_p^q
  =\sqrt{(m\!-\!p\!+\!1)\mu}\ \delta_p^q\ ,
\end{eqnarray}
and similarly
\begin{equation}
  [Z_2]_p^{\ \ q}=\sqrt{p\mu}\ \delta_{p\!+\!1}^q\ .
\end{equation}
These are the known matrix forms of the irreducible blocks, which proves our claim
in (\ref{block-tensor}).

Similar representation can be found for second type blocks of size $(m\!+\!1)\times m$.
One obtains
\begin{equation}
  [Z_{c\!+\!2}]_{a_1a_2\cdots a_m}^{b_1b_2\cdots b_{m\!-\!1}}=
  (m\mu)^{\frac{1}{2}}\delta^{(c}_{a_1}\delta^{b_1}_{a_2}\cdots\delta^{b_{m\!-\!1})}_{a_m}
\end{equation}
for $Z_3$, $Z_4$, after similar calculation.

\section{Mass and supersymmetry of charged particles}\label{mass}

In this appendix, we calculate the masses of bosonic BPS charged particles
which correspond to the open membranes stretched between fractional
M2-branes with the minimal area. We work in the convention of \cite{Lee:2009mm}.
Motivated by the supersymmetry analysis of open M2-branes in appendix \ref{susy-gravity},
we study the solutions to the BPS equation preserving $\xi^{34}=(\xi^{12})^\ast$
part of the supersymmetry. We further impose one of the two projection conditions
$\gamma^0\xi^{12}_\pm=\mp i\xi^{12}_\pm$ for BPS massive particles in their rest frames.
These spinors make the particles $\frac{1}{6}$-BPS. As expected for the smeared open M2's
studied in appendix \ref{susy-gravity}, we find that enhancement of supersymmetry
other than this is impossible. Taking the bosonic fields to be space-independent
in the rest frame, the supersymmetry transformation of four fermions is
\begin{eqnarray}\label{susy-12}
  \delta\Psi^1&\sim&-\gamma^0\xi^{12}D_0Z_2+i\gamma^a\xi^{12}
  \left(A_aZ_2-Z_2\tilde{A}_a\right)+2W^1_{12}\xi^{12}+2W^1_{34}\xi^{34}=0\nonumber\\
  \delta\Psi^2&\sim&\gamma^0\xi^{12}D_0Z_1-i\gamma^a\xi^{12}
  \left(A_aZ_1-Z_1\tilde{A}_a\right)+2W^2_{12}\xi^{12}+2W^2_{34}\xi^{34}=0\nonumber\\
  \delta\Psi^3&\sim&-\gamma^0\xi^{34}D_0Z_4+i\gamma^a\xi^{34}
  \left(A_aZ_4-Z_4\tilde{A}_a\right)+2W^3_{34}\xi^{34}+2W^3_{12}\xi^{12}=0\nonumber\\
  \delta\Psi^4&\sim&\gamma^0\xi^{34}D_0Z_3-i\gamma^a\xi^{34}
  \left(A_aZ_3-Z_3\tilde{A}_a\right)+2W^4_{34}\xi^{34}+2W^4_{34}\xi^{34}=0\ .
\end{eqnarray}
Here, $a\!=\!1,2$ is the spatial vector index. It is easy to see that the last terms
on the right hand sides should be zero separately. Also, the second terms containing
spatial components of the gauge fields vanish separately, as they contain
spinors with different $\gamma^0$ projections.

We first consider the BPS fluctuations connecting a first type block of
size $m$ to another first type block of size $n$. As $\gamma^{012}=1$, we have
$\gamma^{12}\xi^{12}=\pm i\xi^{12}$ and the supersymmetry condition on
the $m\times n$ matrix of $A$ and $(m\!+\!1)\times (n\!+\!1)$ matrix of $\tilde{A}$ is
\begin{equation}\label{vector-SUSY-1}
  A_\pm Z_i=Z_i\tilde{A}_\pm\ \ \ (A_\pm\equiv A_1\pm i A_2)\ ,
\end{equation}
or
\begin{equation}\label{gauge-BPS-11}
  \sqrt{m}[\tilde{A}_\pm]_{a_1a_2\cdots a_m}^{b_1b_2\cdots b_n}=\sqrt{n}
  [A_\pm]_{(a_1\cdots a_{m\!-\!1}}^{(b_1\cdots b_{n\!-\!1}}\delta_{a_m)}^{b_n)}
\end{equation}
in the symmetric $SU(2)$ tensor notation, as summarized in appendix \ref{tensor}.
The choice of sign in $A_\pm$ is correlated with the choice of supersymmetry
projection $\xi^{12}_\pm$.
These gauge fields are further constrained by the Gauss' law,
\begin{equation}\label{gauss}
  \frac{k}{4\pi}\epsilon_{\mu\nu\rho}F^{\nu\rho}=
  i\left(D_\mu Z_\alpha\bar{Z}^\alpha-Z_\alpha D_\mu\bar{Z}^\alpha
  \right)\ ,\ \ \frac{k}{4\pi}\epsilon_{\mu\nu\rho}\tilde{F}^{\nu\rho}=
  i\left(\bar{Z}^\alpha D_\mu Z_\alpha-D_\mu\bar{Z}^\alpha Z_\alpha\right)\ .
\end{equation}
We expand both sides to linear order in the scalars and the gauge fields. The gauge
field would obtain masses of order $\frac{\mu}{k}$ through the Higgs mechanism.
The linearized Gauss' law is given by
\begin{eqnarray}\label{gauss-11-scalar}
  \frac{k}{2\pi}\star dA&=&\mu(m\!+\!n\!+\!2)A-2Z_i\tilde{A}\bar{Z}^i
  +idz_i\bar{Z}^i-iZ_idw^i\nonumber\\
  \frac{k}{2\pi}\star d\tilde{A}&=&-\mu(m\!+\!n)\tilde{A}+2\bar{Z}^iAZ_i+
  i\bar{Z}^idz_i-idw^iZ_i\ ,
\end{eqnarray}
where $z_i$ and $w^i$ are $m\times (n\!+\!1)$ and $(m\!+\!1)\times n$ fluctuations
of $Z_i$ and $\bar{Z}^i$, respectively. The last two terms from scalar fluctuations
are total derivatives, which are absorbed into the linearized gauge
transformations of $A$ and $\tilde{A}$
\begin{equation}
  A\rightarrow A+d\lambda\ ,\ \ \tilde{A}\rightarrow\tilde{A}+d\tilde\lambda\ ,\ \
  z_i\rightarrow z_i+i(\lambda Z_i-Z_i\tilde\lambda)\ ,\ \
  w^i\rightarrow w^i+i(\tilde\lambda\bar{Z}^i-\bar{Z}^i\lambda).
\end{equation}
The two combinations of scalar fluctuations
\begin{equation}\label{gauge-transform}
  i(z_i\bar{Z}^i-Z_iw^i)\ ,\ \ i(\bar{Z}^iz_i-w^iZ_i)
\end{equation}
appearing in (\ref{gauss-11-scalar}) change under the linearized gauge transformation as
\begin{equation}\label{gauge-transform-gauss}
  \hspace*{-1cm}i(z_i\bar{Z}^i-Z_iw^i)\rightarrow i(z_i\bar{Z}^i-Z_iw^i)
  -\mu(m\!+\!n\!+\!2)\lambda+2Z_i\tilde\lambda\bar{Z}^i\ ,\ \
  i(\bar{Z}^iz_i-w^iZ_i)\rightarrow i(\bar{Z}^iz_i-w^iZ_i)
  +\mu(m\!+\!n)\tilde\lambda-2\bar{Z}^i\lambda Z_i\ ,
\end{equation}
and can be set to zero by an appropriate choice of $\lambda,\tilde\lambda$.
This is simply the standard procedure in Chern-Simons-matter theories of letting
the vectors to eat up some scalar degrees of freedom in a Higgs phase. If one advocates
the time-dependent ansatz, this gauge transformation will only affect the time components
of the gauge fields.

In the tensor notation, the Gauss' law equations after the gauge-fixing are given by
\begin{equation}\label{gauss-11}
  \frac{k}{2\pi}\star d\left(\begin{array}{c}
  A_{a_1\cdots a_{m\!-\!1}}^{b_1\cdots b_{n\!-\!1}}\\
  \tilde{A}_{a_1\cdots a_m}^{b_1\cdots b_n}\end{array}\right)=
  \mu\left(\begin{array}{c}(m\!+\!n\!+\!2)A_{a_1\cdots a_{m\!-\!1}}^{b_1\cdots b_{n\!-\!1}}
  -2\sqrt{mn}\tilde{A}_{ca_1\cdots a_{m\!-\!1}}^{cb_1\cdots b_{n\!-\!1}}\\
  -(m\!+\!n)\tilde{A}_{a_1\cdots a_m}^{b_1\cdots b_n}+2\sqrt{mn}A_{(a_1\cdots
  a_{m\!-\!1}}^{(b_1\cdots b_{n\!-\!1}}\delta_{a_m)}^{b_n)}\end{array}\right)\ .
\end{equation}
With our ansatz in which fields are time-dependent only, the time components
of the right hand sides should be all zero as the left hand sides are. This will
constrain the gauge fields as $A_0=0$, $\tilde{A}_0=0$ after the gauge transformation
(\ref{gauge-transform-gauss}) is made. As for the remaining spatial components of the
Gauss' law, the equations can be decomposed into irreducible representations of $SU(2)$, by first
contracting $p$ pairs of $a$ and $b$ type indices and then symmetrizing the remaining
indices. One obtains
\begin{equation}\label{gauss-11-irrep}
  \frac{k}{2\pi}\star d\left(\begin{array}{c}
  A^{(m\!+\!n\!-\!2p)}\\ \tilde{A}^{(m\!+\!n\!-\!2p)}\end{array}\right)=
  \mu\left(\begin{array}{cc}(m\!+\!n\!+\!2)&-2\sqrt{mn}\\ \frac{2p(m\!+\!n\!-\!p\!+\!1)}
  {\sqrt{mn}} &-(m\!+\!n)\end{array}\right)\left(\begin{array}{c}
  A^{(m\!+\!n\!-\!2p)}\\ \tilde{A}^{(m\!+\!n\!-\!2p)}\end{array}\right)\ ,
\end{equation}
where $A$, $\tilde{A}$ are understood to have spatial components only, while $d$
has time component only. The superscripts in parentheses denote the number
of symmetrized $SU(2)$ spinor indices.
This expression is valid for $1\leq p\leq\min(m,n)$. At $p\!=\!0$, the first
line of this equation is void and we only have
\begin{equation}\label{symm-gauge}
  \frac{k}{2\pi}\star d\tilde{A}^{(m\!+\!n)}=-\mu(m\!+\!n)\tilde{A}^{(m\!+\!n)}
\end{equation}
with mass given by $\frac{2\pi\mu(m\!+\!n)}{k}$. In other cases, the matrix
appearing on the right hand side has the following eigenvalues:
\begin{equation}
  -(m\!+\!n\!-\!2p)\ ,\ \ m\!+\!n\!+\!2\!-\!2p.
\end{equation}
As $p$ takes its maximal value $\min(m,n)$, the absolute value of the eigenvalue
has the minimum of $|m\!-\!n|$ or $|m\!-\!n\!+\!2|$ in each case.

So far we have not used the supersymmetry condition on these gauge fields.
Applying the supersymmetry condition (\ref{gauge-BPS-11}), one first observes that
the totally symmetric $\tilde{A}^{(m\!+\!n)}_\pm$ in (\ref{symm-gauge}) is zero,
implying that this mode is non-BPS. The other mode $\tilde{A}_\mp^{(m\!+\!n)}$
may appear to be BPS with the mass $m\!+\!n$ as this does not appear in (\ref{symm-gauge}),
but this is not true. Repeating the analysis of (\ref{gauge-BPS-11}) with $m$ and $n$
flipped (i.e. for the complex conjugate modes), one finds that the
$(n\!+\!1)\times(m\!+\!1)$ matrix of $\tilde{A}_\pm^{(m\!+\!n)}$ is forbidden by
supersymmetry. The mode $\tilde{A}_\mp^{(m\!+\!n)}$ which seem to survive the
condition (\ref{vector-SUSY-1}) is conjugate to this, which form a set
of creation-annihilation operators after quantization using the Chern-Simons term as
the symplectic 2-form. So we find no BPS modes here.

Then turning to the other $\tilde{A}^{(m\!+\!n\!-\!2p)}$
modes with $p\geq 1$, the supersymmetry constrains
\begin{equation}
  \sqrt{\frac{m}{n}}\tilde{A}^{(m\!+\!n\!-\!2p)}_\pm=
  \frac{k(m\!+\!n\!-\!p\!+\!1)}{mn}A^{(m\!+\!n\!-\!2p)}_\pm\ .
\end{equation}
Inserting this back to (\ref{gauss-11-irrep}), one obtains
\begin{equation}
  \frac{k}{2\pi}\star dA_\pm^{(m\!+\!n\!-\!2p)}=\mu\left(
  m\!+\!n\!+\!2\!-\!\frac{2p(m\!+\!n\!-\!p\!+\!1)}{m}\right)A_\pm^{(m\!+\!n\!-\!2p)},\
  \frac{k}{2\pi}\star d\tilde{A}_\pm^{(m\!+\!n\!-\!2p)}=\mu\left(
  m\!-\!n\right)\tilde{A}^{(m\!+\!n\!-\!2p)}_\pm\ ,
\end{equation}
which are compatible equations only when $p\!=\!m$, which is allowed only if $m\leq n$.
Thus, the BPS modes with mass $\frac{2\pi\mu|m\!-\!n|}{k}$ exist when $m<n$. For $m>n$,
the BPS mode with the same mass comes from the conjugate $n\times m$ block of $A$ and
$(n\!+\!1)\times(m\!+\!1)$ block of $\tilde{A}$. The modes $A_\mp$ and $\tilde{A}_\mp$
seem to be all unconstrained by supersymmetry, but they again form pairs of
non-supersymmetric oscillators unless the masses are $\frac{2\pi|m\!-\!n|}{k}$.
So the only allowed vectors preserving $\xi^{12}$ supersymmetry are those with mass $\frac{2\pi|m\!-\!n|}{k}$.

One can also consider the gauge field $A_\mu$ in $m\times(n\!+\!1)$ matrix,
and $\tilde{A}_\mu$ in $(m\!+\!1)\times n$ matrix, connecting the first type
block of size $m$ and the second type block of size $n$. The supersymmetry
condition for the spatial component of gauge fields is
\begin{equation}
  A_\mp Z_m=0\ ,\ \ Z_i\tilde{A}_\pm=0\ \ \ \ (i\!=\!1,2,\ m\!=\!3,4)\ ,
\end{equation}
implying that $A_\mp\!=\!0$, $\tilde{A}_\pm\!=\!0$ for supersymmetry: only
$A_\pm$ and $\tilde{A}_\mp$ are allowed for supersymmetry. The conjugates of the
last unconstrained modes come with $A_\mp$ in $(n\!+\!1)\times m$ and $\tilde{A}_\pm$
in $n\times(m\!+\!1)$, which one can easily show are unconstained from the supersymmetry
condition, as they should be. These modes are the BPS modes.

The linearized Gauss' law for the spatial components is
\begin{equation}\label{gauss-12-scalar}
  \frac{k}{2\pi}\star dA=\mu(m\!+\!n\!+\!1)A+idz_m\bar{Z}^m-iZ_idw^i\ ,\ \
  \frac{k}{2\pi}\star d\tilde{A}=-\mu(m\!+\!n\!+\!1)\tilde{A}+i\bar{Z}^idz_i
  -idw^mZ_m\ ,
\end{equation}
where $z_i$, $z_m$ are $m\times n$ matrices for the fluctuations of
$Z_i$, $Z_m$, respectively, and $w^i$, $w^m$ are $(m\!+\!1)\times(n\!+\!1)$
matrices for the fluctuations of $\bar{Z}^i$, $\bar{Z}^m$, respectively.
Again one can use the linearized gauge transformation to set the scalar
combinations appearing in the Gauss' law to be zero. After that,
one obtains
\begin{equation}\label{gauss-12}
  \frac{k}{2\pi}\star dA=\mu(m\!+\!n\!+\!1)A\ ,\ \
  \frac{k}{2\pi}\star d\tilde{A}=-\mu(m\!+\!n\!+\!1)\tilde{A}\ ,
\end{equation}
all with masses $\frac{2\pi\mu}{k}(m\!+\!n\!+\!1)$. They are BPS
for the modes mentioned in the previous paragraph.

We also consider scalar BPS equations, and again start from the modes
connecting two vacuum blocks of first type. The fact that some scalar degrees of
freedom are constrained by (\ref{gauge-transform-gauss}) will be imposed later,
but it eventually kills all the massless modes.
After expanding the BPS equation to linear order in $z,w$, one obtains
\begin{eqnarray}
  \dot{z}_i&=&\mp\frac{2\pi i}{k}\left[\mu(m\!-n)z_i+z_j(\bar{Z}^jZ_i)-(Z_i\bar{Z}^j)z_j
  +Z_jw^jZ_i-Z_iw^jZ_j\right]\nonumber\\
  \dot{w}^i&=&\pm\frac{2\pi i}{k}\left[\mu(n\!-\!m)w^i+(\bar{Z}^iZ_j)w^j-w^j(Z_j\bar{Z}^i)
  +\bar{Z}^iz_j\bar{Z}^j-\bar{Z}^jz_j\bar{Z}^i\right]\ .
\end{eqnarray}
Using the symmetric tensor notation of appendix \ref{tensor},
these equations become
\begin{eqnarray}\label{BPS-bosonic}
  \left[\dot{z}_i\right]_{a_1\cdots a_{m\!-\!1}}^{b_1\cdots b_n}&=&\mp\frac{2\pi\mu i}{k}
  \left[\frac{}{}\!\right.(m\!-\!n)[z_i]_{a_1\cdots a_{m\!-\!1}}^{b_1\cdots b_n}
  +n[z_j]_{a_1\cdots a_{m\!-\!1}}^{j(b_1\cdots b_{n\!-\!1}}\delta_i^{b_n)}
  -m[z_{(i}]_{a_1\cdots a_{m\!-\!1})}^{b_1\cdots b_n}\nonumber\\
  &&\hspace{4cm}+\sqrt{mn}\left([w^j]_{ja_1\cdots a_{m\!-\!1}}^{(b_1\cdots b_{n\!-\!1}}
  \delta_i^{b_n)}-[w^{(b_1}]_{ia_1\cdots a_{m\!-\!1}}^{b_2\cdots b_n)}
  \right)\left.\frac{}{}\!\right]\nonumber\\
  \left[\dot{w}^i\right]_{a_1\cdots a_m}^{b_1\cdots b_{n\!-\!1}}&=&\pm\frac{2\pi\mu i}{k}
  \left[\frac{}{}\!\right.(n\!-\!m)[w^i]_{a_1\cdots a_m}^{b_1\cdots b_{n\!-\!1}}
  +m[w^j]_{j(a_1\cdots a_{m\!-\!1}}^{b_1\cdots b_{n\!-\!1}}\delta_{a_m)}^i
  -n[w^{(i}]_{a_1\cdots a_m}^{b_1\cdots b_{n\!-\!1})}\nonumber\\
  &&\hspace{4cm}+\sqrt{mn}\left([z_j]_{(a_1\cdots a_{m\!-\!1}}^{jb_1\cdots b_{n\!-\!1}}
  \delta^i_{a_m)}-[z_{(a_1}]_{a_2\cdots a_m}^{ib_2\cdots b_{n\!-\!1})}
  \right)\left.\frac{}{}\!\right]\ .
\end{eqnarray}
We would like to solve these equations after decomposing them to
various irreducible representations of $SU(2)$. Firstly, $z$ appearing in the left
hand side of the first equation is in the product representation of spin
$\frac{1}{2}$ ($i$ index), $\frac{m\!-\!1}{2}$ ($a$ indices) and $\frac{n}{2}$
($b$ indices) representations. Irreducible representations can be formed by suitably
contracting (or anti-symmetrizing) and symmetrizing the indices. We would like
to first (anti-)symmetrize $i$ with another type (say, $a$ type) indices and
then (anti-)symmetrize with the last type of indices.

For the $z$ tensor, the first class of representations contracts $i$ with $b$ type
indices, and then further contract $p\!-\!1$ of the $a$ and $b$ type indices.
The remaining free indices are then symmetrized. In other words, one contracts
$p$ of the upper/lower indices including $i$, and then symmetrize the remaining
$m\!+\!n\!-\!2p$ indices. We call this rank $m\!+\!n\!-\!2p$ tensor
\begin{equation}
  z_A^{(m\!+\!n\!-\!2p)}\sim[z_{c_1}]_{c_2\cdots c_pa_p\cdots a_{m\!-\!1}}^{
  c_1\cdots c_pb_{p\!+\!1}\cdots b_n}
\end{equation}
where it is understood that the uncontracted $a$ and $b$ type indices are symmetrized
(after raising the $a$ type indices with $SU(2)$ invariant tensor $\epsilon^{ab}$). The
subscript $A$ stands for anti-symmetrization of the doublet index $i$. As will be
clear later, this representation is meaningful only for $p\geq 1$.
The second class of representations for the $z$ tensor first raise the $i$ index,
symmetrize with all the $b$ type indices, and then contract $p$ of the upper/lower
indices. The free indices are finally symmetrized. This tensor becomes a linear
combination
\begin{eqnarray}
  [z_j]_{c_1\cdots c_pa_{p\!+\!1}\cdots a_{m\!-\!1}}^{
  (c_1\cdots c_pb_{p\!+\!1}\cdots b_n}\epsilon^{i)j}&=&\frac{n\!-\!p\!+\!1}{n\!+\!1}
  \epsilon^{ij}[z_j]_{c_1\cdots c_pa_{p\!+\!1}\cdots a_{m\!-\!1}}^{
  c_1\cdots c_pb_{p\!+\!1}\cdots b_n}+\frac{p}{n\!+\!1}
  \epsilon^{c_pj}[z_j]_{c_1\cdots c_pa_{p\!+\!1}\cdots a_{m\!-\!1}}^{
  c_1\cdots c_{p\!-\!1}ib_{p\!+\!1}\cdots b_n}\nonumber\\
  &\equiv&\frac{n\!-\!p\!+\!1}{n\!+\!1}z_S^{(m\!+\!n\!-\!2p)}+\frac{p}{n\!+\!1}
  \tilde{z}_A^{(m\!+\!n\!-\!2p)}
\end{eqnarray}
for the `generic' case, and the symmetrization of free indices are understood.
For instance, the first tensor in the last expression is given by contracting
$p$ upper/lower indices in $z$ excluding $i$, and then symmetrizing the remaining
$m\!+\!n\!-\!2p$ including the uncontracted $i$:
\begin{equation}
  z_S^{(m\!+\!n\!-\!2p)}\sim[z_i]_{c_1\cdots c_pa_{p\!+\!1}\cdots a_{m\!-\!1}}^{
  c_1\cdots c_pb_{p\!+\!1}\cdots b_n}
\end{equation}
where the uncontracted indices including $i$ are all understood to be symmetrized.
$S$ stands for symmetrization of the index $i$. For $z_A$,
$p$ is constrained as $1\leq p\leq\min(m,n)$, while in the symmetric case one finds
$0\leq p\leq\min(m\!-\!1,n)$. The tensor on the left hand side exists for
$0\leq p\leq n\!+\!1$. On the other hand, the first tensor on the right hand side
exists only for $0\leq p\leq n$ while the second tensor exists for $1\leq p\leq n\!+\!1$.
In the 'non-generic' cases with $p\!=\!0,n\!+\!1$, one of the two terms on the right
hand side is understood to be absent.

$z_S$, $z_A$, $\tilde{z}_A$ are not independent generally.
For instance, for $m\!=\!2$, $n\!=\!1$, one finds at $p\!=\!1$
\begin{equation}
  [z_A]_a=[z_c]^c_a=\left(\begin{array}{c}(z_1)^1_1+(z_2)^2_1\\
  (z_1)^1_2+(z_2)^2_2\end{array}\right)
  ,\ [z_S]_i=[z_i]_c^c=\left(\begin{array}{c}(z_1)^1_1+(z_1)^2_2\\
  (z_2)^1_1+(z_2)^2_2\end{array}\right),\
  [\tilde{z}_A]^b=\epsilon^{ai}[z_i]_a^b=\left(\begin{array}{c}(z_2)^1_1-(z_1)^1_2\\
  (z_2)^2_1-(z_1)^2_2\end{array}\right).
\end{equation}
From this, one obtains
\begin{equation}
  [z_A]_a=[z_S]_a+\epsilon_{ab}[\tilde{z}_A]^b\ \ \ (\epsilon_{ab}\equiv\epsilon^{ab}).
\end{equation}
Such a relation should be generic, as long as all three tensors are well defined.
In particular, for $p\neq 0,n\!+\!1$ where $z_A$, $z_S$ are all meaningful, one may
take these two to be independent basis for the tensor $z$. For $p\!=\!0$ when $z_A$ does
not exist, $z_S$ totally symmetrizes $i,b$ and $a$, which is the only possible tensor
in this sector. When $p\!=\!n\!+\!1$, the only possible tensor is $\tilde{z}_A$. As the
tensor rank is $m\!+\!n\!-\!2p\!=\!m\!-\!n\!-\!2$, the case with
$p\!=\!n\!+\!1$ exists only for $m\geq n\!+\!2$.

Similarly, $w$ on the left hand side of the
second equation of (\ref{BPS-bosonic}) can be decomposed as
\begin{eqnarray}
  w_A^{(m\!+\!n\!-\!2p)}&\sim&\left[w^{c_1}\right]_{c_1\cdots c_pa_{p\!+\!1}\cdots a_m
  }^{c_2\cdots c_pb_p\cdots b_{n\!-\!1}}\ \ \ (1\leq p\leq\min(m,n))\nonumber\\
  w_S^{(m\!+\!n\!-\!2p)}&\sim&\left[w^i\right]_{c_1\cdots c_pa_{p\!+\!1}\cdots a_m
  }^{c_1\cdots c_pb_{p\!+\!1}\cdots b_{n\!-\!1}}\ \ \ (0\leq p\leq\min(m,n\!-\!1))\ ,
\end{eqnarray}
again with symmetrization of uncontracted indices understood. $w_A$ is again
absent for $p\!=\!0$. For $p\!=\!m\!+\!1$, one has to separately consider
\begin{equation}
  \tilde{w}_A^{m\!+\!n\!-\!2p}\sim\epsilon_{c_pj}
  [w^j]_{c_1\cdots c_{p\!-\!1}a_{p}\cdots a_m}^{
  c_1\cdots c_{p}b_{p\!+\!1}\cdots b_{n\!-\!1}}
\end{equation}
as the only possible tensor, which exists only when $n\geq m\!+\!2$,
like $\tilde{z}_A$.

The equations can be decomposed using the above basis as
\begin{eqnarray}
  \hspace*{-1cm}\dot{z}_A^{(m\!+\!n\!-\!2p)}&=&\mp\frac{2\pi\mu i}{k}\left[
  (m\!-\!p\!+\!1)z_A^{(m\!+\!n\!-\!2p)}\!-\!(m\!-\!p)z_S^{(m\!+\!n\!-\!2p)}
  \!+\!\sqrt{\frac{m}{n}}\left((n\!-\!p\!+\!1)w_A^{(m\!+\!n\!-\!2p)}\!-\!
  (n\!-\!p)w_S^{(m\!+\!n\!-\!2p)}\right)\right]\nonumber\\
  \hspace*{-1cm}\dot{w}_A^{(m\!+\!n\!-\!2p)}&=&\pm\frac{2\pi\mu i}{k}\left[
  (n\!-\!p\!+\!1)w_A^{(m\!+\!n\!-\!2p)}\!-\!(n\!-\!p)w_S^{(m\!+\!n\!-\!2p)}
  \!+\!\sqrt{\frac{n}{m}}\left((m\!-\!p\!+\!1)z_A^{(m\!+\!n\!-\!2p)}\!-\!
  (m\!-\!p)z_S^{(m\!+\!n\!-\!2p)}\right)\right]\nonumber\\
  \hspace*{-1cm}\dot{z}_S^{(m\!+\!n\!-\!2p)}&=&\mp\frac{2\pi\mu i}{k}\left[
  -(n\!-\!p)z_S^{(m\!+\!n\!-\!2p)}-\sqrt{\frac{m}{n}}(n\!-\!p)w_S^{(m\!+\!n\!-\!2p)}
  \right]\nonumber\\
  \hspace*{-1cm}\dot{w}_S^{(m\!+\!n\!-\!2p)}&=&\pm\frac{2\pi\mu i}{k}\left[
  -(m\!-\!p)w_S^{(m\!+\!n\!-\!2p)}-\sqrt{\frac{n}{m}}(m\!-\!p)z_S^{(m\!+\!n\!-\!2p)}
  \right]\ .
\end{eqnarray}
The first and second equations are ignored for $p\!=\!0$.
In these equations, the maximum of $p$ is either $\min(m,n)$, $\min(m\!-\!1,n)$ or
$\min(m,n\!-\!1)$. In the generic case with $p\!\neq\!0$, the $4\times 4$ matrix
\begin{equation}
  \mp\frac{2\pi\mu}{k}\left(\begin{array}{cccc}
  m\!-\!p\!+\!1&\sqrt{\frac{m}{n}}(n\!-\!p\!+\!1)&-(m\!-\!p)&-\sqrt{\frac{m}{n}}
  (n\!-\!p)\\-\sqrt{\frac{n}{m}}(m\!-\!p\!+\!1)&-(n\!-\!p\!+\!1)&\sqrt{\frac{n}{m}}
  (m\!-\!p)&n\!-\!p\\0&0&-(n\!-\!p)&-\sqrt{\frac{m}{n}}(n\!-\!p)\\
  0&0&\sqrt{\frac{n}{m}}(m\!-\!p)&m\!-\!p\end{array}\right)
\end{equation}
appearing in the right hand side would give the BPS masses for various modes as
its eigenvalues. The eigenvalues are two $0$'s and two
$\mp(\pm)\frac{2\pi\mu|m\!-\!n|}{k}$'s, where the $(\pm)$ sign is for $m\gtrless n$.
When $p\!=\!0$, we only consider the $2\times 2$ matrix from the last two equations,
\begin{equation}
  \mp\frac{2\pi\mu}{k}\left(\begin{array}{cc}-n&-\sqrt{mn}\\\sqrt{mn}&m\end{array}\right)
\end{equation}
which mixes $z_S^{(m\!+\!n)}$ and $w_S^{(m\!+\!n)}$. The eigenvalues are
$0$ and $\mp\frac{2\pi\mu(m\!-\!n)}{k}$ (for both $m\gtrless n$). Finally, we
consider the exceptional tensors $\tilde{z}_S$, $\tilde{w}_A$ for $p\!=\!n\!+\!1$
and $p\!=\!m\!+\!1$, respectively. These tensors cannot mix with any other tensors.
They have masses $\propto(m-n)$ and exist only when $m\geq n\!+\!2$ or
$m\leq n\!-\!2$.

It is also easy to study the BPS charged excitations from $Z_3$, $Z_4$ connecting
the vacuum blocks made of $Z_1,Z_2$. One always finds the following BPS equation
for the $m\times(n\!+\!1)$ matrix part
of the fluctuations for $Z_3,Z_4$, connecting two vacuum blocks of first type
and sizes $m$, $n$:
\begin{equation}
  \dot{Z}_{3,4}=\mp\frac{2\pi i\mu(m\!-\!n)}{k}Z_{3,4}\ .
\end{equation}
This gives the mass $\frac{2\pi\mu|m\!-\!n|}{k}$, same as the excitations
from $Z_1,Z_2$. One should further impose
\begin{equation}
  0=W^m_{12}\sim Z_1\bar{Z}^mZ_2-Z_2\bar{Z}^mZ_1
\end{equation}
for $m\!=\!3,4$. We simply count the number of degrees surviving this constraint.
As the number of equations is $n(m\!+\!1)$ for the $m(n\!+\!1)$ variables
for each scalar, one obtains $m\!-\!n$ variables surviving the constraint for $m\!>\!n$.
For $m\!<\!n$, one should consider the $n\times(m\!+\!1)$ blocks to obtain $n\!-\!m$
degrees, after an analysis similar to above. The surviving modes should be in
the $|m\!-\!n|$ dimensional representation of $SU(2)_1$.

The masses of scalars connecting two vacuum blocks of second type can be
studied in exactly the same manner as the analysis above.

We finally consider the massive scalar modes connecting a vacuum block
of first type to another block of second type. Denoting by $z_i$ the
$m\times n$ fluctuation of $Z_i$ and $w^m$ the $(m\!+\!1)\times(n\!+\!1)$
fluctuation of $\bar{Z}^m$, one obtains
\begin{eqnarray}
  [\dot{z}_i]_{(m\!-\!1)}^{(n\!-\!1)}&=&\mp\frac{2\pi\mu i}{k}\left[
  (m\!+\!n\!+\!1)[z_i]_{(m\!-\!1)}^{(n\!-\!1)}-m[z_{(i}]_{(m\!-\!1))}^{(n\!-\!1)}
  +\sqrt{mn}[w^\ast]_{(m\!-\!1)}^{\ast(n\!-\!1)}\right]\nonumber\\
  \left[\dot{w}^m\right]_{(m)}^{(n)}&=&\mp\frac{2\pi\mu i}{k}\left[
  (m\!+\!n\!+\!1)[w^m]_{(m)}^{(n)}-n[w^{\ast}]_{(m)}^{\ast(n}\delta_i^{)}
  +\sqrt{mn}[z_{(}]_{m)}^{\ast(n}\delta_i^{)}\right]
\end{eqnarray}
where the indices written by $\ast$ imply contractions.
The lower/upper indices are for different $SU(2)_1$ and $SU(2)_2$, respectively. We
can either totally symmetrize each type of indices, or anti-symmetrize one pair and then symmetrize
the remainder if there is an extra $i$ or $m$ doublet index. Doing so, one obtains
\begin{eqnarray}
  [\dot{z}_S]_{(m)}^{(n\!-\!1)}&=&\mp\frac{2\pi\mu i}{k}\left[(n\!+\!1)
  [z_S]_{(m)}^{(n\!-\!1)}+\sqrt{mn}[w_A]_{(m)}^{(n\!-\!1)}\right]\nonumber\\
  \left[\dot{w}_A\right]_{(m)}^{(n\!-\!1)}&=&\mp\frac{2\pi\mu i}{k}\left[m[w_A
  ]_{(m)}^{(n\!-\!1)}+\sqrt{\frac{m}{n}}(n\!+\!1)[z_S]_{(m)}^{(n\!-\!1)}\right]\nonumber\\
  \left[\dot{z}_A\right]_{(m\!-\!1)}^{(n\!-\!2)}&=&\mp\frac{2\pi\mu i}{k}
  (m\!+\!n\!+\!1)[z_A]_{(m\!-\!1)}^{(n\!-\!2)}\nonumber\\
  \left[\dot{w}_S\right]_{(m)}^{(n\!+\!1)}&=&\mp\frac{2\pi\mu i}{k}
  (m\!+\!n\!+\!1)[w_S]_{(m)}^{(n\!+\!1)}\ .
\end{eqnarray}
The first two equations contain a $2\times 2$ matrix
\begin{equation}
  \left(\begin{array}{cc}n\!+\!1&\sqrt{mn}\\ \sqrt{\frac{m}{n}}(n\!+\!1)&m\end{array}\right)
\end{equation}
with eigenvalues $0$ and $m\!+\!n\!+\!1$. All modes have mass
$M_{BPS}\!=\!\frac{2\pi\mu(m\!+\!n\!+\!1)}{k}$.
Similar BPS modes come from $m\times n$ block $w^i$ of $\bar{Z}^i$ and
$(m\!+\!1)\times(n\!+\!1)$ block $z_m$ of $Z_m$.

\end{document}